\documentclass[prx,twocolumn,balancelastpage,superscriptaddress,floatfix,longbibliography,nofootinbib,aps,10pt]{revtex4-2}

\usepackage{stylesheet}
\usepackage{comment}
\newcommand{\expect}[1]{\langle #1\rangle}

\begin{document}

\title{Superresolution in Quantum Noise Spectroscopy via Filter Design}

\author{Joseph~T.~Iosue}
\email{jtiosue@gmail.edu}
\affiliation{\JHUAPL}
\affiliation{\QUICS}
\affiliation{\JQI}

\author{Paraj~Titum}
\affiliation{\JHUAPL}
\affiliation{\JHU}

\author{Taohan Lin}
\affiliation{\JHUAPL}

\author{Clare~Lau}
\affiliation{\JHUAPL}

\author{Leigh~M.~Norris}
\altaffiliation{Present address: Quantinuum, Broomfield, Colorado, USA}
\affiliation{\JHUAPL}

\date{February 11, 2026}

\begin{abstract}
    Resolving signals with closely spaced frequencies is central to applications in communications, spectroscopy and sensing. Recent results have shown that quantum sensing protocols can exhibit \emph{superresolution}, the ability to discriminate between spectral lines with arbitrarily small frequency separation. Here, we revisit this problem from the perspective of quantum control theory, utilizing the filter function formalism to derive general, analytic conditions on quantum control protocols for achieving superresolution. Building on these conditions, we develop an optimal control framework, the utility of which is demonstrated through numerical identification of superresolution control protocols in the presence of realistic, experimentally-relevant constraints. We further extend our results to entangled initial states and assess their potential advantage. Our approach is broadly applicable to a wide variety of quantum sensing platforms, and it provides a systematic path to discover novel protocols that surpass conventional resolution limits in these systems.
\end{abstract}

\maketitle


\section{Introduction}
\subsection{Context and Motivation}
Quantum sensors offer the possibility of unrivaled sensitivity in applications such as time keeping, electromagnetic field sensing and gravity sensing~\cite{Degen2017Quantum-sensing,Crawford2021Quantum-Sensing,Aslam2023Quantum-sensors,Bass2024Quantum-sensing,Ye2024Essay:-Quantum-,Patel2025Review-of-qubit,Montenegro2025Review:-Quantum}. Various experimental platforms show promise as quantum sensors, including entangled photons~\cite{AasiLIGO2013}, neutral atoms~\cite{Allred2002,Cox2018}, ions~\cite{Gilmore2017,Gilmore2021}, and solid-state spins~\cite{Taylor2008,Jiang2023,Wang2022}. Many quantum sensing applications can be viewed through the lens of parameter estimation, where the quantum state $\rho(\theta)$ describing the sensor depends on an unknown parameter of interest $\theta$~\cite{Holevo2011Probabilistic-a,Braunstein1994,BRAUNSTEIN1996-Annals,BRAUNSTEIN1996-Physlett}.
For example, when estimating the amplitude or frequency of an electric or magnetic field using a quantum sensor, the parameter of interest is encoded in the phase evolution of a quantum state. The achievable precision in this estimation problem is limited by the quantum  Cram\'{e}r-Rao bound which states that the variance of the estimator is bounded from below by inverse of the quantum Fisher information~\cite{Helstrom1969Quantum-detecti,Holevo2011Probabilistic-a,Braunstein1994,BRAUNSTEIN1996-Annals,BRAUNSTEIN1996-Physlett}.

Another key metric in quantum sensing is \emph{resolution}- the minimum detectable difference between two physical parameters. A prominent example is spatial resolution in optical imaging. Consider the paradigmatic problem of estimating the distance between two point sources of light. Conventional direct imaging suffers from Rayleigh's curse, i.e., it is impossible to resolve two point sources of light below the diffraction limit~\cite{born1999principles-of-o}. In the case of incoherent point sources, however, analyses based on the quantum Fisher information show  that knowledge of the ``centroid'' or midpoint between the sources can be leveraged to achieve \emph{superresolution}, the ability to resolve arbitrarily small separations below the diffraction limit ~\cite{Tsang2016Quantum-informa,Nair2016Far-Field-Super,tsang2016quantum-theory-,Nair2016Interferometric,Tsang2019Quantum-limit-t,Tsang2020Resolving-starl}. Related approaches for achieving superresolution with a quantum sensor have been put forward for an analogous problem in spectral estimation, resolving the frequency separation between two closely spaced peaks in the spectral density of a stochastic signal ~\cite{gefen2019overcoming-reso}. This task arises applications such as nuclear magnetic resonance (NMR) spectroscopy for the identification of molecules and chemical structure~\cite{aslam2017nanoscale-nucle} and in
magnetometry for measurement and detection of weak magnetic fields~\cite{magaletti2022a-quantum-radio}. Using conventional techniques based on Fourier transforms, the minimum resolvable frequency separation scales inversely with the observation time. 
Similar to the imaging problem, however, quantum superresolution sensing protocols are capable of resolving arbitrarily small peak separations below the limit set by the inverse of the observation time ~\cite{gefen2019overcoming-reso,Cao2025Overcoming-freq}.

Using a quantum sensor to the estimate spectral properties of a stochastic signal (e.g., peak separation) falls under the broad umbrella of quantum noise spectroscopy~(QNS)~\cite{Faoro2004,Yuge2011,Young2012}. In QNS, the stochastic signal of interest is typically environmental noise and the goal is estimating the power spectral density (PSD) or ``spectrum'' $S(\omega)$.  QNS protocols can accomplish this by utilizing open-loop control to tailor the spectral response of the quantum sensor. Consider, for example, a qubit sensor subject to dephasing noise and control consisting of transverse $\pi$-pulses. In the limit of instantaneous pulses, the dephasing rate of the sensor at time $T$ takes the form $\chi(T)\propto\int_{-\infty}^\infty d\omega F(\omega,T)S(\omega)$, where the filter function (FF) $F(\omega,T)$ is the frequency-domain representation of the control applied to the sensor ~\cite{paz-silva2014general-transfe,green2013arbitrary-quant,Norris2016}. In QNS, control is selected in order to shape the FF in a way that enables features of the spectrum to be inferred from measured quantities on the sensor, such as $\chi(T)$. This approach has been widely employed to characterize noise across a variety of experimental quantum computing platforms~\cite{Bylander2011,Sung2019,Murphy2022, Chan2018,Romach2015,Frey2017}, as well as in quantum sensing applications for oscillating or stochastic signals~\cite{Hirose2012,Titum2021,schultz2022towards}.

In this work, we approach the frequency resolution problem using filter shaping techniques inspired by QNS and related work in optical superresolution ~\cite{Tsang2016Quantum-informa,Nair2016Far-Field-Super,tsang2016quantum-theory-,Nair2016Interferometric,Tsang2019Quantum-limit-t,Tsang2020Resolving-starl}. Our approach builds on Ref.~\cite{gefen2019overcoming-reso}, which derives necessary and sufficient conditions on the state of a quantum sensor for achieving frequency superresolution. Rather than the sensor state, however, our focus is the control applied to the quantum sensor. We determine conditions on the control filter functions that guarantee frequency superresolution - that is, the ability to resolve arbitrarily small frequency separations in finite time. 
Similar to quantum superresolution in optical imaging, these conditions are satisfiable with knowledge of the centroid, which in this case refers to the center frequency between the peaks. Framing frequency resolution as a control problem allows us to determine operational criteria for achieving superresolution with a quantum sensor in practice. Leveraging tools from optimal and robust quantum control, our FF-based approach furthermore provides a framework to optimize the resolution of a quantum sensor subject to realistic control constraints and imperfections, such as environmental noise.  

\subsection{Summary and main results}
In this work, we consider a stochastic signal that induces dephasing on a controllable qubit sensor. The PSD of the stochastic signal consists of two tones with a known centroid or center frequency. By measuring the sensor, we seek to estimate the separation between the two tones as the frequency spacing becomes arbitrarily small. Details of the two-tone signal, the qubit sensor, and the control setting are described in Sec. \ref{sec:problem}.  In this section, we also introduce necessary background in estimation theory and formalize our definition of frequency superresolution.  The key results of this paper, detailed in the succeeding sections, we summarize below.

\textit{ Superresolution conditions.--}
The first main result of our work, described in \cref{sec:superresolution-conditions}, establishes a set of necessary and sufficient conditions to achieve superresolution with a controlled qubit sensor under the condition that the control is dephasing-preserving. 
The conditions are expressed in terms of the filter function~\cite{kofman2004unified-theory-,cywiifmmode-nelse-nfiski2008how-to-enhance-,paz-silva2014general-transfe,ball2015walsh-synthesiz}, which itself depends \emph{only} on the control and evolution time. 
This highlights a key distinction between our work and that of Ref.~\cite{gefen2019overcoming-reso}; namely, we provide sufficient conditions for superresolution solely in terms of the \textit{control}, whereas Ref.~\cite{gefen2019overcoming-reso} focuses on the state of the quantum sensor.
In \cref{sec:superresolution-conditions}, we also show that the Fisher information (FI), which quantifies the information about the frequency separation contained in measurements of the sensor, is proportional to the second derivative of the filter function at the centroid. This expression allows us to bound the estimation error and derive an upper bound to the maximum achievable FI as well as conditions for optimality of the control protocol.
In \cref{sec:instantaneous-superresolution}, we analyze the performance of two families of control sequences capable of achieving superresolution, namely, free evolution and Carr-Purcell-Meiboom-Gill (CPMG) sequences with instantaneous $\pi$-pulses. 

\textit{Robustness.--}
We examine the robustness of superresolution in the presence of background environmental noise in \cref{sec:noise}. Strictly speaking, superresolution is not robust; in the limit of vanishing frequency separation, an arbitrarily small amount of background noise will cause the FI to vanish. Nevertheless, from the perspective of practical implementation, these protocols can still achieve enhanced resolution. Our analysis shows that the overlap of the filter function with the noise spectrum plays a critical role in determining the performance. Consequently, it is advantageous to design control protocols to minimize this overlap while at the same time satisfying the conditions required for superresolution. As a concrete example, we demonstrate the utility of superresolution protocols through simulations that compare the performance of various control protocols in the presence of realistic noise.

\textit{Optimized continuous control protocols.--}
In \cref{sec:continuous-superresolution}, we go beyond dephasing preserving control and examine continuous control protocols involving off-axis driving that are capable of achieving superresolution.
Through analytic and numerical FF analysis, we find that continuous control protocols can outperform both the free and CPMG superresolution protocols in certain regimes.
A major contribution of our work detailed in \cref{sec:numerical-optimization} is a numerical procedure to determine controls that optimize resolution subject to various constraints and nonidealities, such as environmental noise and finite control pulse duration.
Using this procedure, we identify numerically optimized controls that outperform all the aforementioned protocols in certain regimes.

\textit{Entanglement enhancement.--}
In \cref{sec:entanglement}, we analyze superresolution protocols in which multiple qubit sensors are prepared in an entangled initial state. Overall, protocols leveraging entangled states require fewer total resources in terms of the number of qubit sensors, and the number of measurements. 
Notably, due to the stochastic nature of the signal, the entanglement advantage is a constant factor (independent of the number of qubits entangled) that is a function of the desired relative error. 

\textit{Detailed comparison to classical methods.--}
Finally, in \cref{sec:comparisons}, we compare the performance of our control-based superresolution protocols to traditional quantum estimation methods such as quantum noise spectroscopy.
We also compare the performance of our protocol to information-theoretically optimal classical protocols, such as MUSIC, by analyzing the FI. These comparisons elucidate when control-based superresolution protocols should be used, and when other methods are more favorable.

\section{Background}
\label{sec:problem}
\subsection{Two-tone signal}
Our aim is to use a quantum sensor to estimate the frequency separation between two tones in the PSD of a stochastic signal $\gamma(t)$. We assume that $\gamma(t)$ is Gaussian, wide-sense stationary and zero-mean. If \( \angles*{\cdot} \) denotes the ensemble average, the mean and autocorrelation of  \( \gamma(t) \) are given by \( \angles{\gamma(t)} \coloneqq 0 \) and  \( C(\tau)\coloneqq \angles{\gamma(t)\gamma(t+\tau)} \), respectively. Since \( \gamma(t) \) is Gaussian and zero-mean, its statistical properties are completely determined by $C(\tau)$ or, equivalently, by its PSD,
\begin{equation}
    \label{eq:spectral-density}
    S_\gamma(\omega)
    \coloneqq \int_{-\infty}^{\infty} \e^{-\i\omega\tau} C(\tau) \dd\tau.
\end{equation}
Here, \( \omega \) denotes angular frequency. 

Motivated by Ref.~\cite{gefen2019overcoming-reso}, we consider an idealized, NMR-relevant signal with the following form,
\begin{equation}
    \label{eq:signal}
    \begin{aligned}
        \gamma(t) = g \big[ & A_1 \cos(\omega_1 t) + A_2 \cos(\omega_2 t)          \\
                            & + B_1 \sin(\omega_1 t) + B_2 \sin(\omega_2 t) \big].
    \end{aligned}
\end{equation}
Here, $\omega_1$ and $\omega_2$ are the tones, $g$ is a constant with units of angular frequency,  and \( A_1,A_2,B_1 \) and \( B_2 \) are i.i.d.~normal random variables with zero mean and unit variance, satisfying $\expect{A_iA_j}=\expect{B_iB_j}=\delta_{ij}$ and $\expect{A_iA_j}=0$. The spectral density of $\gamma(t)$, determined via \eqref{eq:spectral-density}, is
\begin{equation}
    \label{eq:signal-S-omega}
    \begin{aligned}
        S_\gamma(\omega) = \pi g^2 \big[
         & \delta(\omega_1-\omega) + \delta(\omega_1+\omega)   \\
         & + \delta(\omega_2-\omega) + \delta(\omega_2+\omega)
            \big].
    \end{aligned}
\end{equation}
Since the spectral density is an even function satisfying $S_\gamma(\omega)=S_\gamma(-\omega)$, we restrict to positive frequencies. For $\omega>0$, the spectral density is a sum of two Dirac delta functions centered at the tones $\omega_1$ and $\omega_2$.
We seek to estimate the frequency separation between the tones, \( \Delta\omega \coloneqq \omega_2 - \omega_1 \), up to a specified relative error  \( \delta\)
or absolute error \( \delta \Delta\omega \).
Specifically, we are interested in the limiting case \( \Delta\omega\to 0 \), where conventional techniques based on Fourier transforms fail to resolve the tone separation in finite time. We assume that the centroid \( \omega_c\coloneqq (\omega_1+\omega_2)/2 \) is known, as it can be estimated with established techniques ~\cite{gefen2019overcoming-reso}. 

\subsection{Superresolution: definition}
In the context of metrology, estimating $\Delta\omega$ falls  within the framework of parameter estimation.
Accordingly, the Fisher information provides a natural metric for quantifying the information about $\Delta\omega$ that can be extracted from a measured quantity on the sensor $\bm{\xi}$. If \( P(\xi \vert \Delta\omega) \) is the conditional probability of measuring outcome $\bm\xi=\xi$ for a given frequency separation \( \Delta\omega \), the Fisher information is defined as~\cite{Holevo2011Probabilistic-a,Helstrom1969Quantum-detecti},
\begin{equation}
    \label{eq:fisher-information-definition}
    \FI_{\bm\xi}({\Delta\omega}) \coloneqq \sum_{\xi} \frac{1}{P(\xi \vert \Delta\omega)} \parentheses{\frac{\partial P(\xi \vert \Delta\omega)}{\partial \Delta\omega}}^2 .
\end{equation}
The Fisher information $\FI_{\bm\xi}({\Delta\omega})$ sets the fundamental lower bound on the variance of an unbiased estimator $\widetilde{\Delta\omega}$ that depends on measurements of $\bm\xi$~\cite{Holevo2011Probabilistic-a,Helstrom1969Quantum-detecti}. Given \( N \) measurements of $\bm\xi$, the Cram\'{e}r-Rao bound is  \(\textrm{var}\pargs*{\widetilde{\Delta\omega}} >\frac{1}{N \FI_{\bm\xi}({\Delta\omega})} \). In this setting, we now formally define superresolution in terms of the Fisher information. An estimation protocol based on measurements of an observable $\bm\xi$ exhibits \emph{superresolution} if the associated Fisher information satisfies
\begin{equation}
    \lim_{\Delta\omega\to 0} \FI_{\bm\xi}({\Delta\omega}) > 0 .
    \label{eq:superres-condition}
\end{equation}
Additionally, we require that sensor evolves for a finite time, even in the limit \( \Delta\omega\to 0 \). The optimal superresolution protocol can be identified by maximizing \(\FI_{\bm\xi}({\Delta\omega})\).

\subsection{Estimation with quantum sensor}
\label{sec:model}

We are interested in \emph{quantum} protocols for for estimating $\Delta\omega$. Specifically, we seek to identify conditions for a quantum sensor evolving in the presence of the signal and control to exhibit superresolution. We consider a prototypical model of a controllable quantum sensor represented by a noisy qubit that is coupled longitudinally to the signal of interest,
\begin{equation}
    \label{eq:hamiltonian}
    H(t) = \gamma(t)\sigma^z + c(t) \sigma^x,
\end{equation}
where \( \sigma^z \) and \( \sigma^x \) are the Pauli operators and  $c(t)$ is a control waveform. 
We take the quantization axis of the sensor to be along $\sigma^z$, so that the dynamics are pure dephasing in absence of control.

The quantum estimation protocol consists of three steps, (i)~prepare the sensor in specified initial state, (ii) let the sensor evolve in the presence of the signal and control for a total time \( T = \kappa\tau \), with \( \tau\coloneqq 2\pi/\omega_c \), and (iii) measure the sensor in a specified basis. In this work we will fix the initial state of the qubit sensor to be \( \ket+=\frac{1}{\sqrt{2}}(\ket0+\ket1) \) and the measurement basis to be $\{\ket{+},\ket{-}\}$; in \cref{ap:robustness}, we show that this can be done without loss of generality. 
Thus, the problem of quantum superresolution amounts to identifying \( \kappa \) and a functional form of the control \( c(t) \) that maximize \( \lim_{\Delta\omega\to 0} \FI_{\psi}(\Delta\omega) \), where $\psi\in\{+,-\}$ denotes the outcome of a measurement on the sensor.  

We consider two types of control commonly utilized in quantum sensing:
\begin{itemize}
    \item [(i)] \emph{Instantaneous $\pi$-pulse control}---\ This control consists of sequences $\pi$-pulses where the interpulse spacing is much longer than the duration of the pulses. In this regime, the $\pi$-pulses can be approximated as instantaneous and the control waveform
    is a sum of Dirac delta functions, \( c(t) = \frac{\pi}{2}\sum_{i=1}^M \delta(t-t_i) \). This control is also single-axis or dephasing-preserving, as detailed in Sec. \ref{sec:superresolution-conditions}.
    \item [(ii)] \emph {Continuous controls}---\ In this case, the control waveform $c(t)$ has an arbitrary functional dependence. The resulting dynamics are multiaxis and do not preserve dephasing.
\end{itemize}
Instantaneous control is an idealized control scheme that can exactly meet the superresolution condition in \cref{eq:superres-condition}. As we discuss in \cref{sec:instantaneous-superresolution,sec:continuous-superresolution}, continuous control is more appealing from the perspective of optimizing resolution in the presence of imperfections, such as environmental noise.

\section{Superresolution under dephasing-preserving control}
\label{sec:superresolution-conditions}
First, we analyze superresolution in the form of instantaneous $\pi$-pulses, which preserves the dephasing dynamics of the sensor. Note that the superresolution conditions discussed in this section are specific to the signal spectrum defined in \cref{eq:signal-S-omega}, though the procedure we present can be straightforwardly adapted to other signal spectra. If the $\pi$-pulses are applied at times $\{t_1,\ldots, t_M\}$, the sensor Hamiltonian in Eq. \eqref{eq:hamiltonian} becomes
\begin{equation}
    H(t) = \gamma(t)\sigma^z + \frac{\pi}{2}\sum_{i=1}^M \delta(t-t_i) \sigma^x.
\end{equation}
To isolate the dynamical contribution of the signal, we transform into the ``toggling frame'' or interaction picture associated with the control. Sensor dynamics in the toggling frame are generated by the Hamiltonian
\begin{align}\label{eq::toggling}
\tilde{H}(t)= f(t) \gamma(t)\sigma^z,
\end{align}
where $f(t)$ is a control switching function that changes sign from $\pm 1$ to $\mp 1$ each time a $\pi$-pulse is applied. Note that instantaneous $\pi$-pulses are dephasing preserving in the sense that $\tilde{H}(t)$ acts solely along $\sigma_z$ and generates pure dephasing with a rate dependent on $\gamma(t)$ and $f(t)$.

\subsection{Conditions for superresolution and FI bound}
We establish the conditions for superresolution under dephasing-preserving control by examining the FI as $\Delta\omega\rightarrow 0$.
To assess the FI, we first determine the conditional probabilities $P(\pm|\Delta\omega)$, which have exact analytic solutions for dephasing-preserving control and Gaussian $\gamma(t)$. If $\expect{\cdot}$ denotes the ensemble average over all realizations of the two-tone signal $\gamma(t)$ and $\tilde{U}(T)=e^{-i\int_0^Tds\tilde{H}(s)}$ is the toggling-frame propagator, the conditional probabilities are given by
\begin{align}
\label{eq:probability-inst}
&P(+|\Delta\omega)=\big\langle\big|\bra{+}\tilde{U}(T)\ket{+}\big|^2\big\rangle= \frac{1}{2}  +\frac{1}{2}e^{-\chi(T) },\\
\label{eq:probability-inst-minus}
&P(-|\Delta\omega)=1-P(+|\Delta\omega).
\end{align}
The decay parameter associated with the dephasing of the sensor is
$\chi(T)=\frac{2}{\pi}\int_{0}^{\infty} \dd\omega\ S_{\gamma}(\omega) F(\omega,T)$, 
where $F(\omega, T) = |\int_0^T f(t) \e^{\i \omega t} \dd t|^2$ is the FF or frequency-domain representation of the control. By Parseval's theorem, the filter function satisfies the normalization condition $\int_{-\infty}^\infty \dd \omega F(\omega,T)=2\pi\int_0^Tdt|f(t)|^2=2\pi T$. 

Given the spectrum of the signal \( \gamma(t) \) in \cref{eq:signal-S-omega}, the decay parameter becomes
\begin{align}
\chi(T)=&\,2g^2\big[F(\omega_c-\Delta\omega/2, T)+F(\omega_c+\Delta\omega/2, T)\big].\label{eq::chi}
\end{align}
From the conditional probabilities in Eqs. (\ref{eq:probability-inst}) and (\ref{eq:probability-inst-minus}), the FI is given by
\begin{widetext}
\begin{align}
    \label{eq:fisher-information-qprotocol}
    \FI_{\psi}(\Delta\omega)  = &\,\frac{1}{2}\parentheses{\coth\chi(T)-1}\parentheses{\frac{\partial \chi(T)}{\partial \Delta\omega} }^2\\
=&\begin{cases}
            g^2 F''(\omega_c, T) + \bigO{\Delta\omega^2}                                                                               & \text{if } F(\omega_c, T) = 0,    \\
            \frac{1}{2} g^4 F''(\omega_c, T)^2 \brackets{ \coth\pargs{4g^2 F(\omega_c, T)} - 1 }\Delta\omega^2 + \bigO{\Delta\omega^4} & \text{if } F(\omega_c, T) \neq 0,
        \end{cases} 
    \end{align}
\end{widetext}
where in the last lines we have expanded $\chi(T)$ about $\Delta\omega\sim 0$. Note that the absence of a term linear in $\Delta\omega$ is due to the fact that the PSD, FF and, thus, $\chi(T)$ are symmetric under \( \Delta\omega\to -\Delta\omega \). 
Observe that when $F(\omega_c, T) = 0 $,  the leading order term in the FI is independent of $\Delta\omega$, implying that 
\begin{align}\label{eq::FI_limit}
\text{lim}_{\Delta\omega\rightarrow 0}\FI_{\psi}(\Delta\omega)=
g^2 F''(\omega_c, T).
\end{align}
Thus, the FI remains finite as $\Delta\omega\rightarrow 0$ so long as $F''(\omega_c, T)>0$. Because $F(\omega,T)>0$, observe that $F''(\omega_c, T)$ is necessarily nonnegative if $F'(\omega_c, T)=0$.
Thus, a control protocol with associated FF $F(\omega, T)$ and total evolution time \( T \) exhibits superresolution if \( F(\omega_c,T) = 0 \) and \( F''(\omega_c, T) > 0 \). In other words, to achieve superresolution, the filter function at the centroid must vanish and have a non-vanishing second derivative. 

The optimal control for superresolution maximizes the FI as $\Delta\omega\rightarrow 0$. Building off of Ref.~\cite{Pang2017Optimal-adaptiv}, Ref.~\cite{gefen2019overcoming-reso} proved an upper bound on the Fisher information, \(\text{lim}_{\Delta\omega\rightarrow 0} \FI_{\psi}(\Delta\omega) \leq \frac{16g^2 T^4}{\pi^2} \). 
Using our control-based superresolution conditions, we tighten this bound to
\begin{align}
    \FI_{\psi}(\Delta\omega) \leq \frac{g^2 T^4}{6}.
    \label{eq:Fi-bound}
\end{align}
To prove this, observe that the second derivative of the FF is bounded by
\begin{align}
     F''(\omega, T) &= \frac{d^2}{d\omega^2}\int_0^T \int_0^T f(t_1)f(t_2) \e^{\i\omega (t_1-t_2)} \dd t_1  \dd t_2, \nonumber \\
     &\leq \int_0^T \int_0^T (t_1-t_2)^2 \dd t_1 \dd t_2 =\frac{T^4}{6},
\end{align}
where in the second line we replaced the integrand by its absolute value. Substituting this expression into the FI in Eq. \eqref{eq::FI_limit} establishes an upper bound for any control protocol whose FF satisfies the superresolution conditions.

\subsection{Estimation and error analysis}
\label{sec:estimation-error-analysis}

To estimate $\Delta\omega$, we repeat the estimation protocol described in Sec. \ref{sec:model} a total of $N$ times and compute the  survival probability.  If $N_+$ is the number of times the sensor was measured in the $\ket{+}$ state, the measured survival probability is $\widetilde{P}=N_+/N$. Note that the expected value of $\widetilde{P}$ is equivalent to $P(+|\Delta\omega)$. Using Eqs. \eqref{eq:probability-inst} and \eqref{eq::chi}, we can write the expected survival probability in the limit of small $\Delta\omega$ as
\begin{align}\label{eq::Pexpand}
    \expect{\widetilde{P}} \equiv P(+|\Delta\omega)= a - b \Delta\omega^2 + \bigO{\Delta\omega^4},
\end{align}
where
\begin{align}\label{eq::a}
&a = \frac{1}{2}[1+e^{-4g^2F(\omega_c,T)}],\\\label{eq::b}
&b = \frac{g^2}{4}e^{-4g^2F(\omega_c,T)} F''(\omega_c,T). 
\end{align}
From the measured survival probability, we can construct an estimator of the frequency separation via
\( \widetilde{\Delta\omega} \equiv \sqrt{(a - \widetilde P)/b} \).
Note that the superresolution condition $F(\omega_c, T)=0$ implies $a=1$. If we were instead to use the survival probability of the $\ket{-}$ state, $1-P$, observe that 
superresolution is equivalently achieved when $a=0$.
This matches the criterion described in Ref.~\cite{gefen2019overcoming-reso}, where they find \( \angles*{P} \propto \Delta\omega^2 \) for small \( \Delta\omega^2 \) as a condition for superresolution.

Using error propagation, we find that to achieve an estimation error of \( \sim \varepsilon \) in \( \Delta\omega \), the survival probability must be measured with an error less than \( 2 b \Delta\omega \varepsilon \).
From Chebyshev's inequality, we find
\begin{equation}
    \begin{aligned}
         & \Pr{}{\abs{\widetilde{P} - \expect{\widetilde{P}}} \geq 2 b \Delta\omega \varepsilon}                                              \\
         & \qquad \leq \frac{a(1-a) + \abs{b(1-2a)}\Delta\omega^2}{4N b^2\Delta\omega^2 \varepsilon^2} + \bigO{\Delta\omega^4},
    \end{aligned}
\end{equation}
where we have used the binomial distribution of $N_+$ and $\text{var}[\widetilde{P}]=\text{var}[N_+]/N^2=$ \( \expect{\widetilde{P}}(1- \expect{\widetilde{P}})/N \).
Hence, in order to achieve a fixed absolute error \( \varepsilon \) in our estimate of \( \Delta\omega \), the number of measurements must scale as
\begin{equation}
    \label{eq:number-measurements}
    N \sim
    \begin{cases}
        \frac{1}{4 b \varepsilon^2}                      & \text{if } a \in \set*{0, 1} \\
        \frac{a(1-a)}{4b^2 \Delta\omega^2 \varepsilon^2} & \text{if } 0 < a < 1.
    \end{cases}
\end{equation} Note this scaling  matches the scaling determined from the Cram\'er-Rao bound, which implies \( N \geq 1/(\varepsilon^2 \FI_{\Delta\omega}) \).

It is more natural to work in terms of relative error \( \delta = \varepsilon / \Delta\omega \) since we are considering the limit as \( \Delta\omega\to 0 \).
Hence, to achieve a fixed relative accuracy \( \delta \) in the superresolution case (\( a\in\set*{0,1} \)), it is sufficient for the number of measurements to scale as
\begin{equation}
    \label{eq:sr-num-measurements}
    N \sim \frac{1}{4 b \delta^2 \Delta\omega^2} = \frac{1}{g^2 F''(\omega_c, \kappa\tau)\delta^2 \Delta\omega^2}.
\end{equation}
If the superresolution condition is not satisfied ($0<a<1$),
one must measure \( N \sim 1/\Delta\omega^4 \) times to achieve the same relative accuracy.

Thus far in our analysis, in order to build intuition, we have assumed the probability of measuring the qubit in $\ket{+}$ is  $P(+|\Delta\omega)$ each time we measure. Recall that $P(+|\Delta\omega)$ was derived by taking the ensemble average over all realizations of $\gamma(t)$. In reality, however, each measurement corresponds to a different realization $\gamma(t)$ specified by the random variables \( A_1,A_2,B_1, \) and \( B_2 \); see \cref{eq:hamiltonian} for the model. Accordingly, for a single shot of the experiment, the probability of measuring the qubit in $\ket{+}$, $P(+|\Delta\omega,\gamma)$, depends on the realization of $\gamma(t)$.

Formally, sampling $N$ times yields $N$ random variables $\psi_1,\dots,\psi_N\in \{0,1\}$,
so that our estimator becomes 
\begin{equation}
    \label{eq:estimator}
    \widetilde{\Delta\omega} = \sqrt{\frac{1}{b}\parentheses{a - \frac{1}{N}\sum_{i=1}^N \psi_i}}.
\end{equation}
The random variable $\psi_i$ can be obtained by first picking $A_1,A_2,B_1,B_2$ i.i.d.~mean zero, unit variance Gaussians for \cref{eq:signal}, evolving the $\ket+$ state with the Hamiltonian in \cref{eq:hamiltonian}, and then measuring in the $\sigma^x$ basis.
A plus measurement corresponds to $\psi_i=1$ while a minus measurement corresponds to $\psi_i=0$.
We denote by $\Expval f(\psi_i)$ the action of performing such the averaging with respect to this procedure.

We are first interested in showing that $\widetilde{\Delta\omega}$ is an unbiased estimator.
In \cref{sec:noiseless-error-analysis}, we show that in the small $\Delta\omega$ limit, 
\begin{equation}
    \frac{\abs*{\Expval \widetilde{\Delta\omega} - \Delta\omega}}{\Delta\omega}
    \leq 
    \frac{a(1-a)}{2b^2 N \Delta\omega^4} + \frac{\abs{2a-1}}{2b N \Delta\omega^2}.
\end{equation}
Thus, in the superresolution case $a\in\{0,1\}$, the estimator is unbiased as $\Delta\omega\to 0$ as long as $N \gtrsim \frac{1}{b\Delta\omega^2}$.

Given this, Chebyshev's inequality can be used to upper bound $\Pr{}{\abs*{\widetilde{\Delta\omega}-\Delta\omega} \geq \delta\Delta\omega}$ in terms of the variance of the estimator.
Then, by simply using 
that for any integer $k>0$, $\Expval \psi_i^k = \Expval \psi_i = a - b\Delta\omega^2$,
we can upper bound the variance.
The result is that in the limit that $\Delta\omega\to0$, in order for our estimator to achieve a relative error of at most $\delta$ with probability at least $1-p$, it suffices to measure $N$ times as long as 
\begin{equation}
    N \gtrsim \frac{1}{bp\delta^2 \Delta\omega^2},
\end{equation}
matching \cref{eq:number-measurements}.

\begin{figure}
    \centering
        \includegraphics[width=.8\columnwidth]{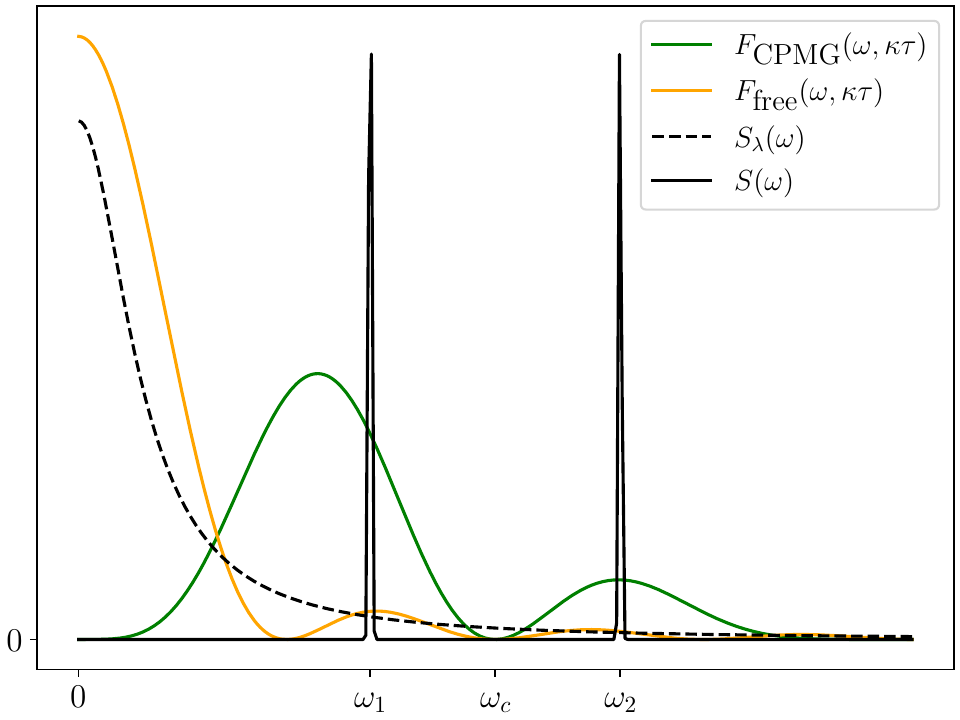}
    \caption{Illustration of the filter functions corresponding to the free evolution and CPMG superresolution protocols (with \(\kappa=2\)), and the (approximate) signal \( S(\omega) \) and noise spectrum $S_\lambda(\omega)$. 
        }
    \label{fig:filter-functions-and-instantaneous-SR-performance}
\end{figure}

\subsection{Superresolution protocols}
\label{sec:instantaneous-superresolution}
We consider two superresolution protocols using dephasing preserving control: (i) free evolution (FE-SR), and (ii) Carr-Purcell-Meiboom-Gill (CPMG-SR), each with a total evolution time of $T=\kappa \tau$. Both FE and CPMG are commonly used in conventional quantum sensing experiments for frequency or phase estimation. In the context of frequency resolution,
FE-SR is discussed in Ref.~\cite{gefen2019overcoming-reso}. While CPMG-SR is not explicitly considered in Ref.~\cite{gefen2019overcoming-reso}, the authors identify a protocol using instantaneous $\pi$-pulses that achieves close to optimal performance (within a constant factor) in terms of the Fisher information. For the purposes of demonstrating our FF-based approach, we restrict to the FE-SR and CPMG-SR protocols, noting that the application of our approach to other dephasing preserving control protocols is straightforward.

In FE-SR, no control is applied to the sensor, so that the corresponding switching function is \( f(t) = 1 \) for all \( t \). In the CPMG-SR protocol, we apply instantaneous $\pi$-pulses separated by an idle time $\tau\equiv 2\pi/\omega_c$. For a CPMG-SR protocol of duration $T=\kappa\tau$ with $\kappa$ an even integer, pulses are applied at times \( \tau/2,3\tau/2,\dots,(2\kappa-1)\tau/2 \). 

A schematic plot of the filter functions generated by FE-SR and CPMG-SR for $\kappa=2$ along with a two-tone spectrum is shown in \cref{fig:filter-functions-and-instantaneous-SR-performance}. Note that the choice of $T=\kappa\tau$ ensures that both filter functions vanish at the centroid, as can be seen from the analytic expression for the filter functions,
\begin{equation}
    \label{eq:free-cpmg-filter-functions}
    \begin{split}
    &\qquad F(\omega,\kappa\tau) = \\
    &\begin{cases}
\frac{4}{\omega^2} \sin^2\pargs*{\frac{\pi \kappa \omega}{\omega_c}} & \text{free (}\kappa\in\bbN) \\
\frac{16}{\omega^2} \sec^2\pargs*{\frac{\pi\omega}{\omega_c}} \sin^2\pargs*{\frac{\pi \kappa \omega}{\omega_c}} \sin^4\pargs*{\frac{\pi\omega}{2\omega_c}}
        & \text{CPMG (}\kappa \in 2\bbN). 
    \end{cases}
    \end{split}
\end{equation}
Furthermore, examining the shape of the filter function near the centroid, it is clear that the CPMG filter function has a larger second-derivative, and thus, greater Fisher information.
From the analytical definition of the filter functions and Eq. \eqref{eq::FI_limit}, the FI in the limit of vanishing $\Delta\omega$ is
\begin{equation}
    \label{eq:free-cpmg-fi}
    \FI_{\Delta\omega\to 0} = \begin{cases}
        \frac{8\pi^2 g^2 \kappa^2}{\omega_c^4},  & \text{free (}\kappa\in\bbN),    \\
        \frac{32\pi^2 g^2 \kappa^2}{\omega_c^4}, & \text{CPMG (}\kappa\in2\bbN) .
    \end{cases}
\end{equation}
Using \cref{eq:sr-num-measurements}, it is clear that in order the achieve a fixed relative estimation error, the CPMG protocol requires \( 4 \) times \emph{fewer} measurements than free evolution.

\begin{figure*}
    \centering
    \begin{minipage}{0.495\textwidth}
        \includegraphics[width=\textwidth]{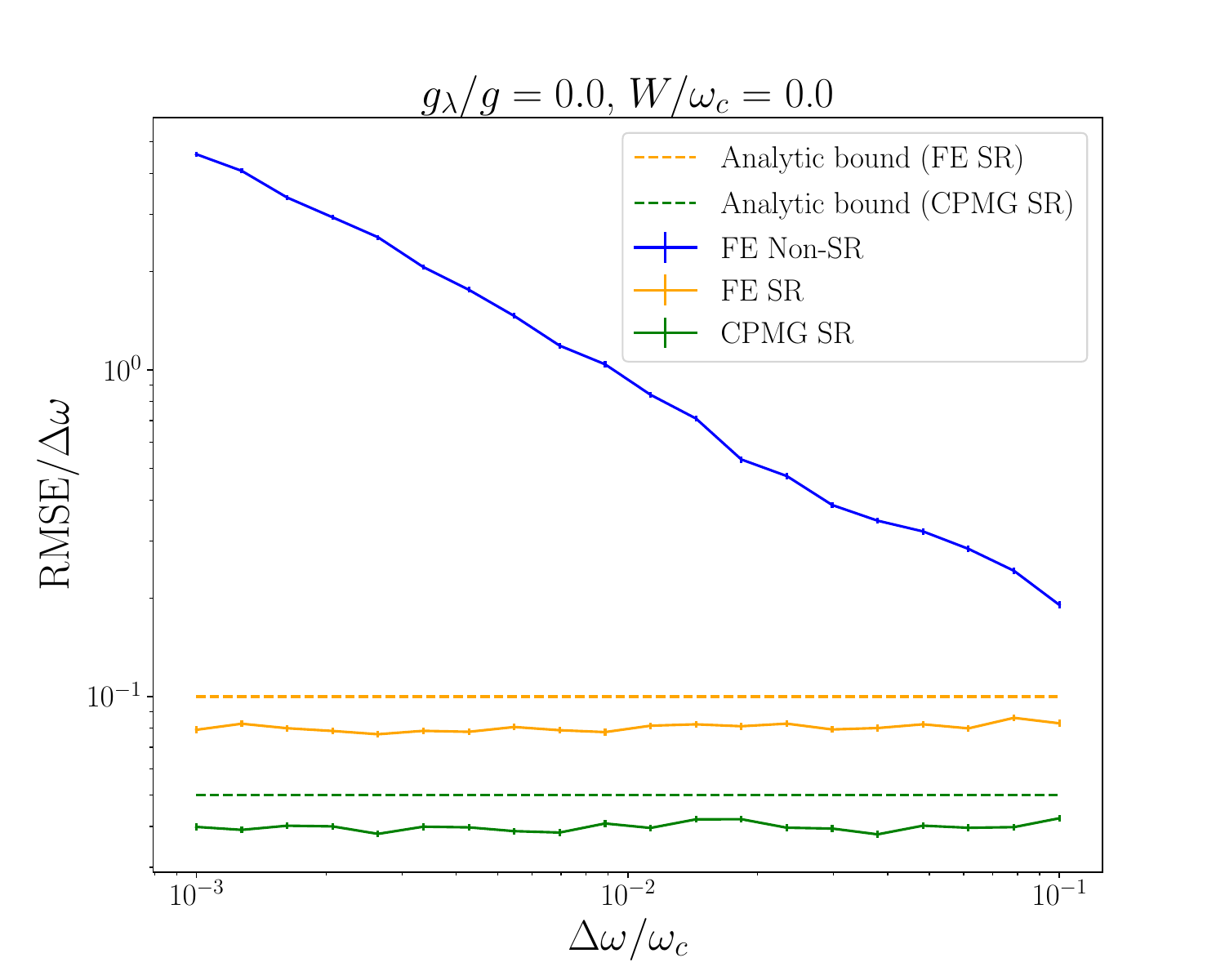}
    \end{minipage}
    \begin{minipage}{0.495\textwidth}
    \includegraphics[width=\textwidth]{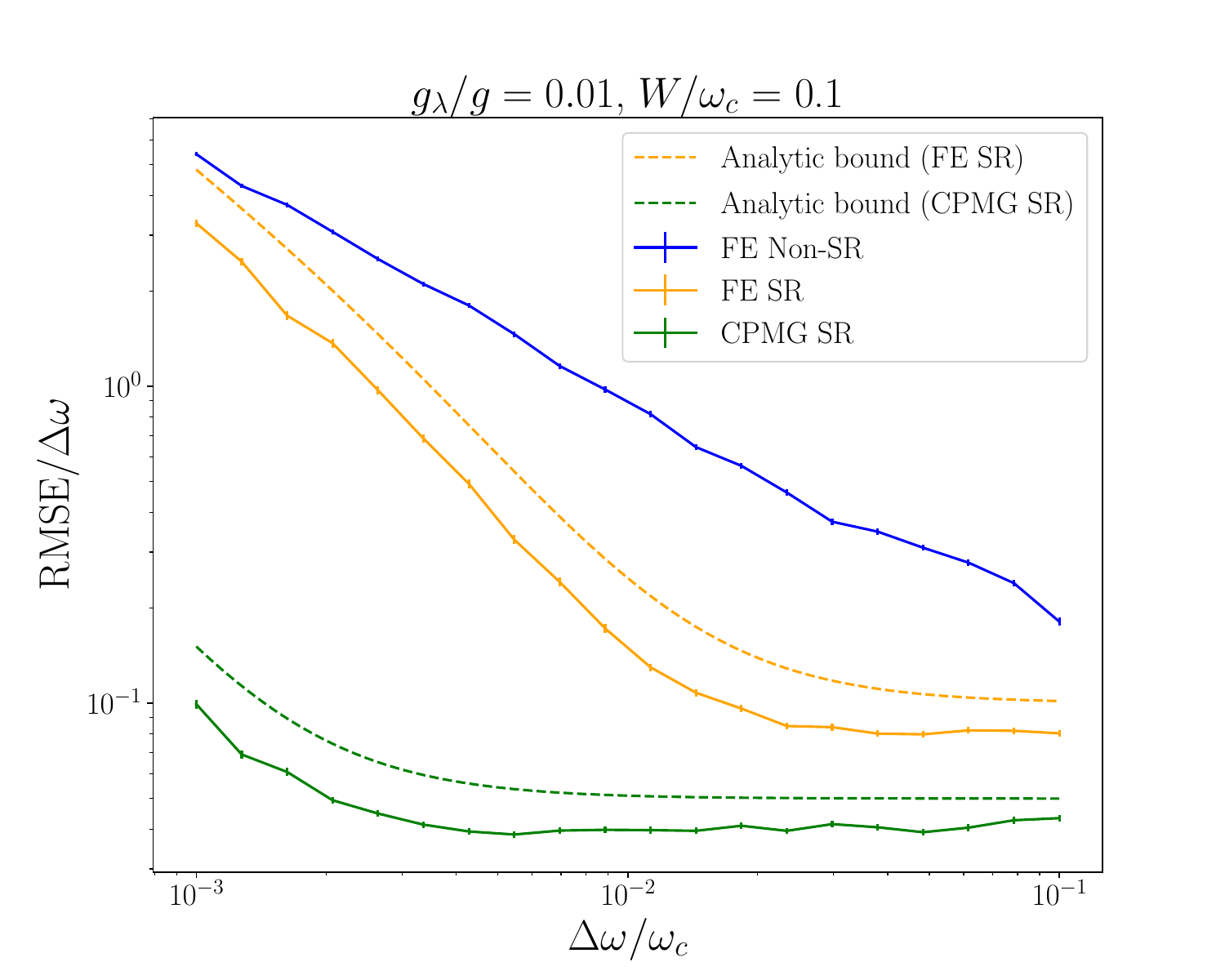}
    \end{minipage}
    \caption{Given the parameters \( g = 0.1 \) MHz, \( \omega_c = 1 \) MHz, and desired relative error \( \delta=0.1 \), we numerically simulate the free evolution protocol with \( \kappa = 5/2 \) which is \emph{not} a superresolution protocol (FE Non-SR), the free evolution superresolution protocol with \( \kappa = 2 \) (FE SR), and the CPMG superresolution protocol with \( \kappa = 2 \) (CPMG SR). For each of the protocols, we utilize \( N_{\rm free} \) samples as given in \cref{eq:number-measurements}, which sets the number of samples for the FE-SR protocol, with $\delta=0.1$ .
        \textbf{(Left)}: Noiseless case; the Lorentzian noise parameters are set to \( g_\lambda = W = 0 \). As \( \Delta\omega\to 0 \), the SR protocols achieve the desired relative accuracy while the non-SR protocol fails.
        \textbf{(Right)}: Noisy case; the Lorentzian noise parameters are set to \( g_\lambda = 0.001 \) MHz and FWHM \( W = 0.1 \) Hz.
        The CPMG superresolution protocol performs significantly better due to the fact that the corresponding filter function overlaps with the noise spectrum significantly less.
        In both figures, the analytic error bounds are from \cref{ap:error-analysis}.}
    \label{fig:MSE-numerics}
\end{figure*}

Next, we numerically compare the performance of FE-SR and CPMG-SR to a dephasing preserving control protocol that does not satisfy the superresolution condition, $F(\omega_c,T)=0$. In \cref{fig:MSE-numerics}~(left), we compare the performance of three protocols with a fixed number of measurements: (i) FE-SR with \( \kappa = 2 \); (ii) ``FE-Non-SR'', free evolution with \( \kappa = 5/2 \); and CPMG-SR  with \( \kappa = 2 \). To compare the protocols on equal footing, we set the number of measurements to be the number needed for FE-SR to achieve a desired relative accuracy of $\delta=0.1$, as given in Eq. \eqref{eq:sr-num-measurements}. As expected from the analytic expressions, CPMG-SR produces an estimate of $\Delta\omega$ with a lower root mean square error (RMSE) than  FE-SR.  Additionally, the RMSE produced by FE-Non-SR is greater than both FE-SR and CPMG-SR, and increases as $\Delta\omega\rightarrow 0$. This is expected, as FE-Non-SR requires a factor of \( \sim \Delta\omega^{-2} \) \emph{more} measurements to achieve the same error as the superresolution protocols, despite the fact that it uses a larger evolution time. 



\subsection{Performance with noise}
\label{sec:noise}
Next, we analyze the performance of superresolution protocols with dephasing preserving control in the prescence of environmental dephasing noise. We model the dephasing noise by adding an additional stationary, Gaussian, zero-mean, stochastic process $\lambda(t)$ to the sensor Hamiltonian in Eq. \eqref{eq:hamiltonian},
\begin{equation}
    \label{eq:noise-hamiltonian}
    H(t) = [\gamma(t)+\lambda(t)]\sigma^z + c(t) \sigma^x.
\end{equation}
Specializing to instantaneous $\pi$-pulses, we find the toggling frame Hamiltonian takes a form similar to Eq. \eqref{eq::toggling},
\begin{align}
\tilde{H}(t)= f(t) [\gamma(t)+\lambda(t)]\sigma^z.
\end{align}
With the added noise, the conditional probability in Eq. \eqref{eq:probability-inst} becomes
\begin{align}
P(+|\Delta\omega)= \frac{1}{2}  +\frac{1}{2}e^{-\chi(T)-\chi_\lambda(T) },
\end{align}
where $\chi_\lambda(T)=\frac{2}{\pi}\int_{0}^{\infty} \dd\omega\ S_{\lambda}(\omega) F(\omega,T)$ and $S_{\lambda}(\omega)$ is the PSD of $\lambda(t)$.
Expanding to leading order in $\Delta\omega$ in the limit of weak signal and noise, the conditional probability is given by
\begin{align}
    P(+|\Delta\omega) = 1-  2g^2F(\omega_c,T) - \frac{1}{2}\chi_\lambda(T) + O(\Delta\omega^2).
\end{align}
Clearly, with non-zero noise, $\lim _{\Delta\omega\rightarrow 0} P(+|\Delta\omega) < 1$, even when the superresolution condition ($F(\omega_c,T)=0$) is satisfied.
It follows that superresolution is not robust to noise, in the sense that \( \lim_{\Delta\omega\to0}\FI = 0 \), as we found in \cref{sec:estimation-error-analysis} that the conditional probability must be $1$ (or $0$) to achieve a non-zero limiting Fisher information.
Nonetheless, for finite but small \( \Delta\omega \), superresolution protocols can still yield performance that is better than non-superresolution protocols.

The simplest scenario is a Markovian model for the noise; this corresponds to a white noise process with $S_\lambda(\omega)=\Gamma$. While this is an idealized model for noise, since the noise power is infinite, it is still instructive for analyzing the estimation error. For this white noise model, $\chi_\lambda=2\Gamma \kappa \tau$, independent of the applied control sequences. Thus, regardless of the control protocol, the noise modifies the expected survival probability in exactly the same way. A key consequence of the noise is that \( \widetilde{\Delta\omega} \equiv \sqrt{(a - \widetilde P)/b} \) is no longer an unbiased estimator of the frequency separation, meaning that $\langle\widetilde{\Delta\omega}\rangle\neq\Delta\omega$.

Assuming $N$ measurements to estimate the probability, and then using the noiseless expressions from \cref{sec:estimation-error-analysis} for the estimation procedure, the estimated value, $\widetilde{\Delta \omega}$, has a bias ($\bias(\widetilde{\Delta \omega})$) in addition to the variance ($\variance(\widetilde{\Delta \omega}))$). Expanding to the lowest order in $\Delta \omega$ and $\gamma$,
\begin{align}
    \bias(\widetilde{\Delta \omega})&= \frac{2 \gamma \omega_c^3}{\alpha\pi g^2\kappa\Delta\omega},\\
    \variance({\widetilde{\Delta \omega}}) &\leq \frac{1}{4 N \alpha \pi^2 g^2 \kappa^2 / \omega_c^4}
\end{align}
where $\alpha=2$ for FE-SR and $\alpha=8$ for CPMG-SR. A detailed derivation is provided in \cref{ap:error-analysis-white-noise}. 
We note here that while the variance matches that of the noiseless case, the relative error scales as $1/\Delta\omega^2$ due to the bias and thus the protocol does not technically achieve superresolution.
The bias in the estimate can be removed by independently characterizing the noise and correcting for its contribution with a new (unbiased) estimator \cite{RiberiPRA2023}. The protocol still does not achieve superresolution, however, because the variance of this new estimator has an additional factor of $1/\Delta\omega^2$, as we show in \cref{sec:subtract-bias}.

Next, let us consider non-Markovian noise, as given by a time-correlated Gaussian noise process. Specifically, we consider a Lorentzian noise spectrum
\begin{equation}
    \label{eq:lorentzian-noise}
    S_\lambda(\omega) = \frac{g_\lambda^2}{\pi} \frac{2W}{4\omega^2 + W^2},
\end{equation}
where \( W \) is the full width at half maximum (FWHM) and \( g_\lambda \) is the strength of the noise. Such a process is useful to model noise in quantum sensors such as nitrogen-vacancy defects in diamonds~\cite{Bar-Gill2012Suppression-of-}. In this case, the applying control can significantly modify $\chi_\lambda$. Since $\chi_\lambda$ is proportional to the overlap integral between the noise spectrum and the filter function, minimizing this overlap reduces $\chi_\lambda$, and consequently, the errors from noise. In \cref{fig:filter-functions-and-instantaneous-SR-performance}, we plot the filter functions for FE-SR and CPMG-SR with \( \kappa=2 \) alongside. 
As can be seen in the figure, the CPMG-SR filter function has as substantially smaller overlap with the noise than does FE-SR, and we therefore expect that the benefit of CPMG-SR over FE-SR will become even more pronounced in the presence of Lorentzian noise (and more generally any low frequency noise).

In fact, in \cref{ap:error-analysis-lorentzian-noise} we perform detailed analysis of CPMG-SR and FE-SR with the addition of Lorentzian noise calculating the bias and variance in the estimator. We find that, given the same number of measurements, CPMG-SR will achieve a relative error of \( \sim \sqrt{\frac{W}{\kappa^2 \omega_c}}  \) times the relative error that FE-SR achieves.
For example, when \( \kappa = 2 \), CPMG-SR achieves a relative error of less than \( \sqrt{\frac{W}{15\omega_c}} \) times the relative error that FE-SR achieves.
Note that as the FWHM of the noise grows, the overlap of the CPMG filter function increases more and more so that at very large FHWM, the advantage is lost.
However, for low frequency noise, we see a large benefit stemming from the filter function having low weight on low frequencies.

The effects discussed above can be seen clearly in numerical simulations shown in \cref{fig:MSE-numerics}.
The simulations are performed by letting all three of the protocols have access to \( N = \frac{\omega_c^4}{8\pi^2 g^2 \kappa^2 \delta^2 \Delta\omega^2} \) measurements, with \( \kappa = 2 \).
From \cref{eq:sr-num-measurements}, it follows that in the noiseless case, FE-SR should achieve a relative error of at most \( \delta \) and CPMG-SR a relative error of at most \( \delta/4 \), while FE-Non-SR should achieve a relative error that grows as \( \Delta\omega \) decreases. This is confirmed numerically in \cref{fig:MSE-numerics}.
Furthermore, \cref{ap:error-analysis-lorentzian-noise} shows that in the noisy case with small FWHM, we see FE-SR initially achieves a relative error of \( \delta \), but the relative error grows as \( \Delta\omega \) decreases due to the noise.
On the other hand, CPMG-SR remains at a relative error of \( \sim \delta/4 \) for much smaller values of \( \Delta\omega \) (due to the \( \sqrt{\frac{W}{15\omega_c}} \) factor derived above), until eventually it also must grow as \( \Delta\omega\to 0 \).

Finally, we mention one peculiarity in \cref{fig:MSE-numerics}.
Recall that each of the methods, including FE-Non-SR, has access to \( N \sim \Delta\omega^{-2} \) measurements.
Therefore, if we are sampling from the probability distribution defined by \cref{eq:probability-inst} for FE-Non-SR, the relative error should scale as \( \Delta\omega^{-2} \) as \( \Delta\omega\to 0 \) in both the noisy and noiseless cases.
However, the protocols are not sampling from \( \angles{P} \), but rather \( P \).
We showed in \cref{sec:estimation-error-analysis} that this distinction is unimportant for superresolution protocols, but the same analysis with a non-superresolution protocol reveals that the distinction \emph{is} important unless \( N \sim \Delta\omega^{-4} \).
This finite sampling error results in estimates \( \widetilde{P} \) of \( \angles{P} \) that can sometimes yield estimates of \( \Delta\omega^2 \) that are negative.
In the numerical simulations presented in \cref{fig:MSE-numerics}, we combat this effect by simply taking the absolute value of the estimate.
Nonetheless, this approach affects the expected scaling, and indeed the FE-Non-SR relative error does not increase exactly as \( \Delta\omega^{-2} \) as \( \Delta\omega\to0 \).

Our analysis of superresolution with non-Markovian noise shows that choosing superresolution controls such that the FF has minimal overlap with the noise PSD can yield dramatic improvements.
It is therefore natural to numerically search for controls that satisfy the superresolution criteria and minimize overlap with the noise. In general, we are interested in controls such that \( F(\omega_c,\kappa\tau) = 0 \), \( F''(\omega_c, \kappa\tau) \) is maximized, and \( \chi_\lambda \) is minimized. These criteria are well-suited for numerical optimization, which we explore in \cref{sec:numerical-optimization}.
In the next section, we generalize our discussion from instantaneous controls to continuous controls; this enables us to utilize standard numerical optimization libraries. Furthermore, we show how such optimization can incorporate constraints and yield substantial benefits over protocols such as CPMG-SR.

\section{Superresolution under multiaxis control}
\label{sec:continuous-superresolution}
We now consider continuous control protocols; these are protocols where we relax the constraint that the control function consists of instantaneous $\pi$-pulses.
In this scenario, as shown in \cref{ap:continuous-filter-function}, the expected survival probability $\expect{\tilde{P}}=P(+|\Delta\omega)$ is 
\begin{widetext}
    \begin{equation}
        \label{eq:continuous-probability}
        \angles*{\tilde{P}} =
        1-g^2\parentheses{2F^{(2)}(\omega_c, \kappa\tau) + \frac{2g^2}{3} F^{(4)}(\omega_c, \omega_c, \kappa\tau)}
        + g^2 \Delta\omega^2 \parentheses{
            \frac{1}{4}F^{(2)\prime\prime}(\omega_c, \kappa\tau) + \frac{g^2}{3} F^{(4)\prime\prime}(\omega_c, \omega_c, \kappa\tau)
        }
        + \bigO{\Delta\omega^4, g^6},
    \end{equation}
\end{widetext}
where the expansion is to leading order in $\Delta\omega$ and the primes denote derivatives with respect to the first argument.
$F^{(s)}$ denotes the $s^{\rm th}$ order FF defined in \cref{ap:continuous-filter-function}.
We see that the superresolution criteria derived in \cref{sec:superresolution-conditions} are still valid with respect to \( F^{(2)} \) as long as \( g^2/\Delta\omega^2 \) is small.
When \( c(t) \) is an instantaneous control, then $F^{(2)} = F$ and $F^{(4)}(\omega_1, \omega_2, t) = 3 F(\omega_1, t) F(\omega_2, t)$, therefore reducing to \cref{eq:probability-inst} as expected.

Due to the higher order terms in the expansion (i.e., $\bigO*{g^4}$ terms), superresolution cannot be achieved exactly. 
However, if the control satisfies the superresolution criteria at lowest order, meaning that, \( F^{(2)}(\omega_c, \kappa\tau) = 0 \) and \( F^{(2)\prime\prime}(\omega_c, \kappa\tau) > 0 \),
then \cref{eq:continuous-probability} implies a region of \( \Delta\omega \) where the corresponding protocol effectively behaves as a superresolution protocol.
In particular, using \cref{eq:continuous-probability} and requiring that the constant term be small compared to the $\bigO{\Delta\omega^2}$ term, the region is given by all values \( \Delta\omega \) satisfying
\begin{equation}
    \label{eq:cc-valid-regime}
    \Delta\omega^2 \gtrsim \frac{8 g^2 F^{(4)}(\omega_c, \omega_c, \kappa\tau)}{3 F^{(2)\prime\prime}(\omega_c, \kappa\tau)}.
\end{equation}
One major benefit of considering continuous controls is the ability to numerical optimize over a continuous space to find particularly good protocols.
Indeed in \cref{sec:numerical-optimization}, we specifically discuss numerically optimizing to find protocols that attempt to maximize this region by minimize \( F^{(4)}(\omega_c, \omega_c, \kappa\tau) \).

\begin{figure*}
    \centering
    \begin{minipage}{.49\textwidth}
        \includegraphics[width=\textwidth]{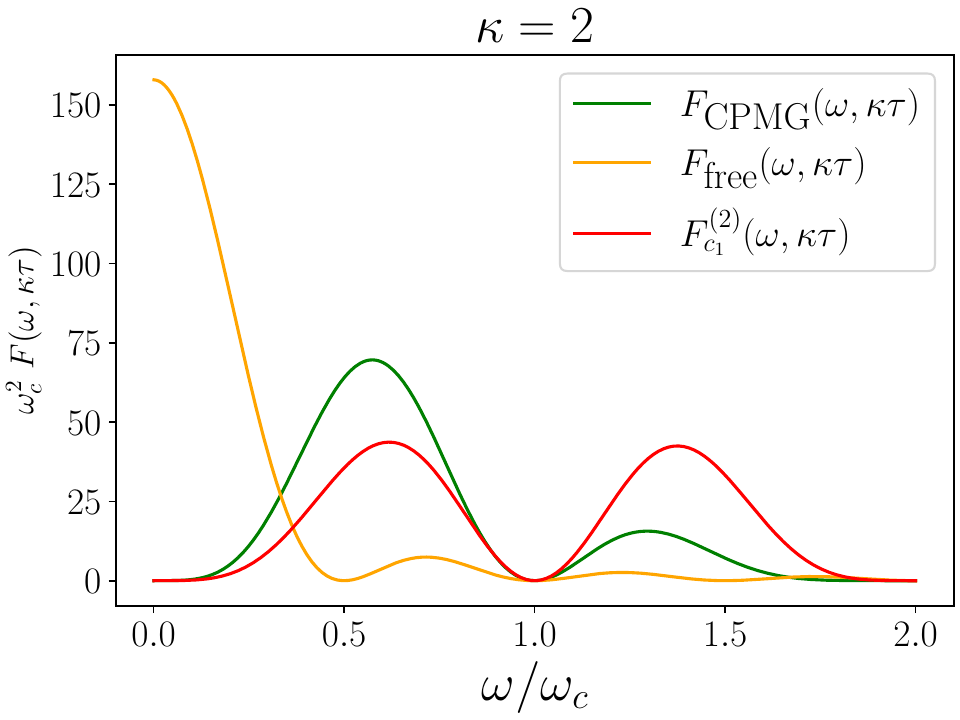}
    \end{minipage}
    \hfill
    \begin{minipage}{.49\textwidth}
        \includegraphics[width=\textwidth]{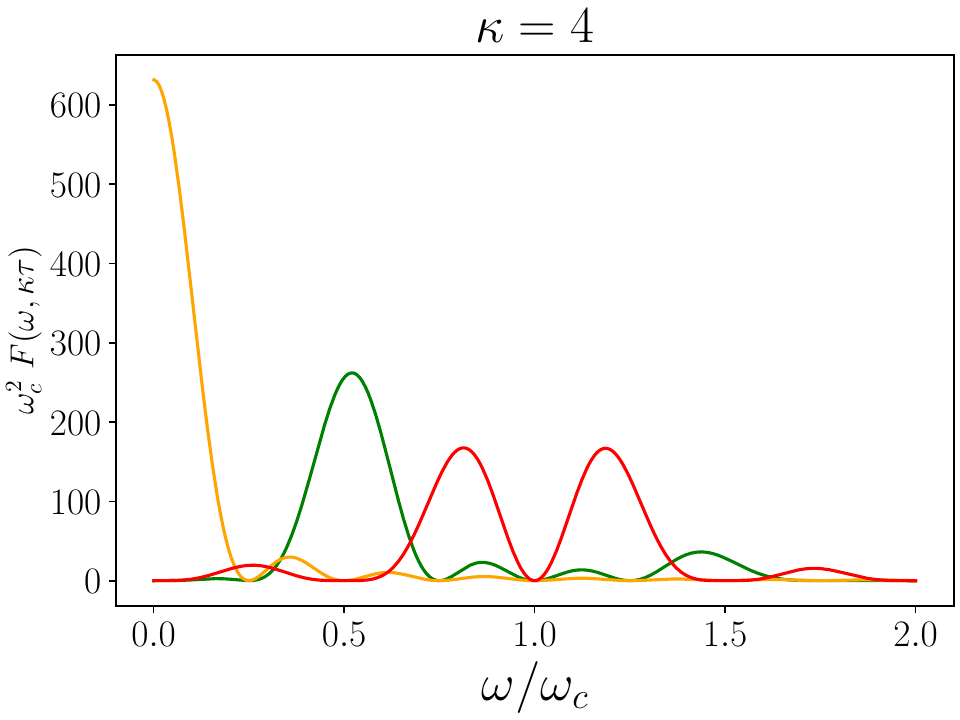}
    \end{minipage}
    \caption{The filter functions for the \( c_1(t) \) protocol (continuous controls) and for the CPMG and free superresolution protocols (instantaneous controls). Note that we are plotting the lowest order filter function. For the instantaneous, dephasing-preserving control protocols, the lowest order filter function is the only one that matters; however, for continuous controls protocols, higher order filter functions become important. (Left) Filter functions for \( \kappa = 2 \). (Right) Filter functions for \( \kappa =4 \).}
    \label{fig:filterfunctions}
\end{figure*}

Before numerically optimizing, though, we analytically study a specific example of continuous control to demonstrate that continuous controls can be beneficial compared to instantaneous, dephasing-preserving control.
We define the control waveform
\begin{equation}
    \label{eq:c1-protocol}
    c_1(t) = -\frac{1}{2} + \frac{\pi}{2} \delta\pargs{t-\frac{\pi\kappa}{\omega_c}},
\end{equation}
which represents a constant drive along $\sigma^x$ with a single instantaneous \( \pi \) pulse halfway through.
When \( \kappa \) is an integer, the filter function \( F^{(2)}_{c_1} \) for \( c_1 \) obeys the superresolution criteria, meaning that $F^{(2)}_{c_1}(\omega_c, \kappa\tau) = 0$ and $F^{(c_1)\prime\prime}(\omega_c, \kappa\tau) > 0$.

\emph{If} the higher filter functions were also to vanish at the centroid, then \( c_1 \) would yield superresolution with the Fisher information
\begin{equation}
    \label{eq:c1-fisher-information}
    \lim_{\Delta\omega\to 0} \FI_{\Delta\omega}^{(c_1)} = \frac{\pi^4 g^2 \kappa^4}{\omega_c^4}
    \qquad (\text{if } \kappa \in \bbN ).
\end{equation}
\cref{eq:cc-valid-regime} then implies that the \( c_1 \) protocol behaves as a superresolution protocol with this Fisher information in the region \( \Delta\omega \gtrsim \sqrt{\frac{3g^2}{2\pi^2}} \sim g/3 \).
We will hence refer to the \( c_1 \) protocol as \( c_1 \)-SR

We can immediately see a benefit of \( c_1 \) as compared to FE-SR and CPMG-SR.
Similar to the instantaneous control protocol derived in Ref.~\cite{gefen2019overcoming-reso}, note that the Fisher information scaling of the \( c_1 \) protocol scales with the fourth power of the total time.
Furthermore, comparing their filter functions shown in \cref{fig:filterfunctions},
\( c_1 \)-SR will be effected less by low frequency noise, especially as \( \kappa \) increases, due to its lower weight on low frequencies.

Indeed, we see this effect in numerical simulations presented in \cref{fig:cc-mse}.
We can explicitly see \( c_1 \)-SR perform better as \( g \) decreases, and the gap between the achieved relative error of CPMG-SR and \( c_1 \)-SR increase as the Lorentzian noise strength increases when \( g \) is small.

\subsection{Numerical optimization of superresolution protocols}
\label{sec:numerical-optimization}

\begin{figure*}
    \centering
    \begin{minipage}{.49\textwidth}
        \includegraphics[width=\textwidth]{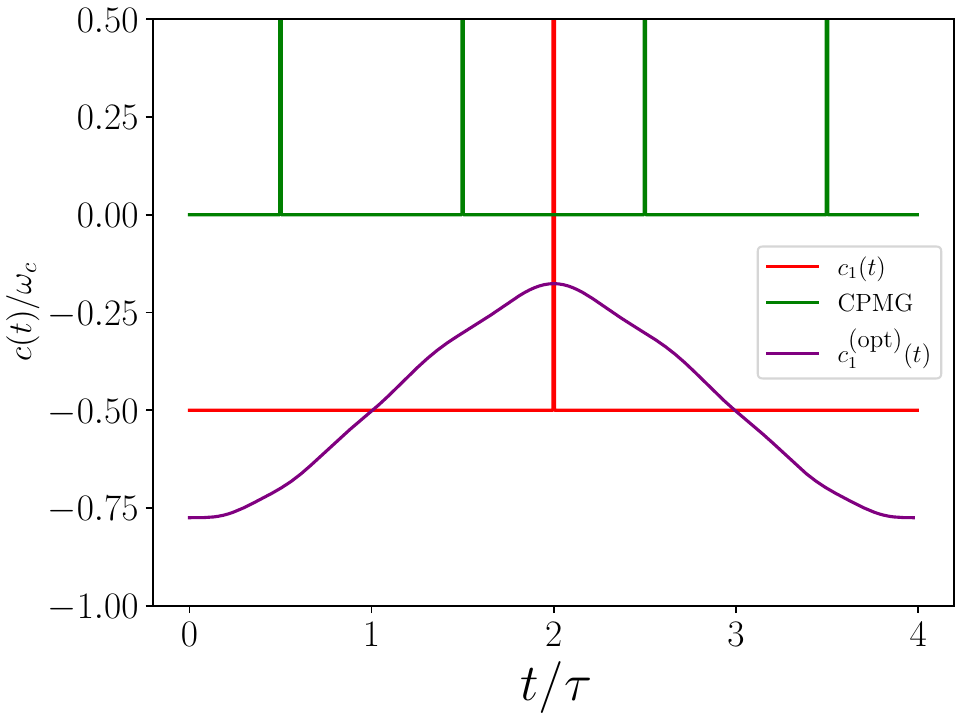}
    \end{minipage}
    \hfill
    \begin{minipage}{.49\textwidth}
        \includegraphics[width=\textwidth]{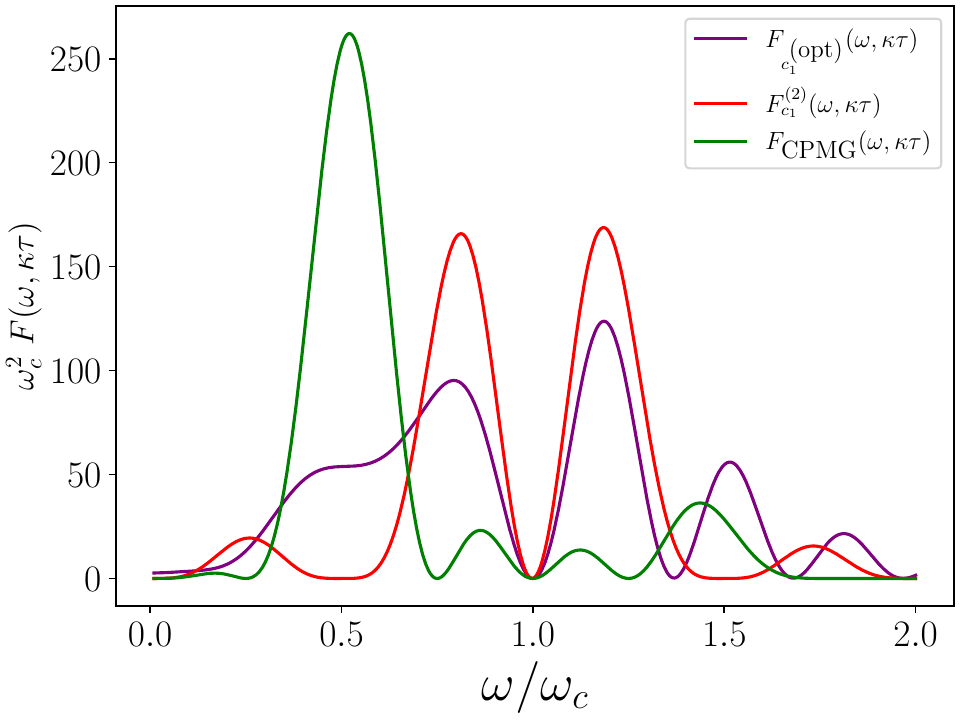}
    \end{minipage}
    \caption{For \( \kappa=4 \). (Left): \( c_1 \), \( c_1^{\rm (opt)} \), and CPMG control sequences. (Right): Corresponding filter functions. The green and red filter functions are the same as in the right of \cref{fig:filterfunctions}.} 
    \label{fig:all-controls}
\end{figure*}

\begin{figure*}
    \centering
    \includegraphics[width=\textwidth]{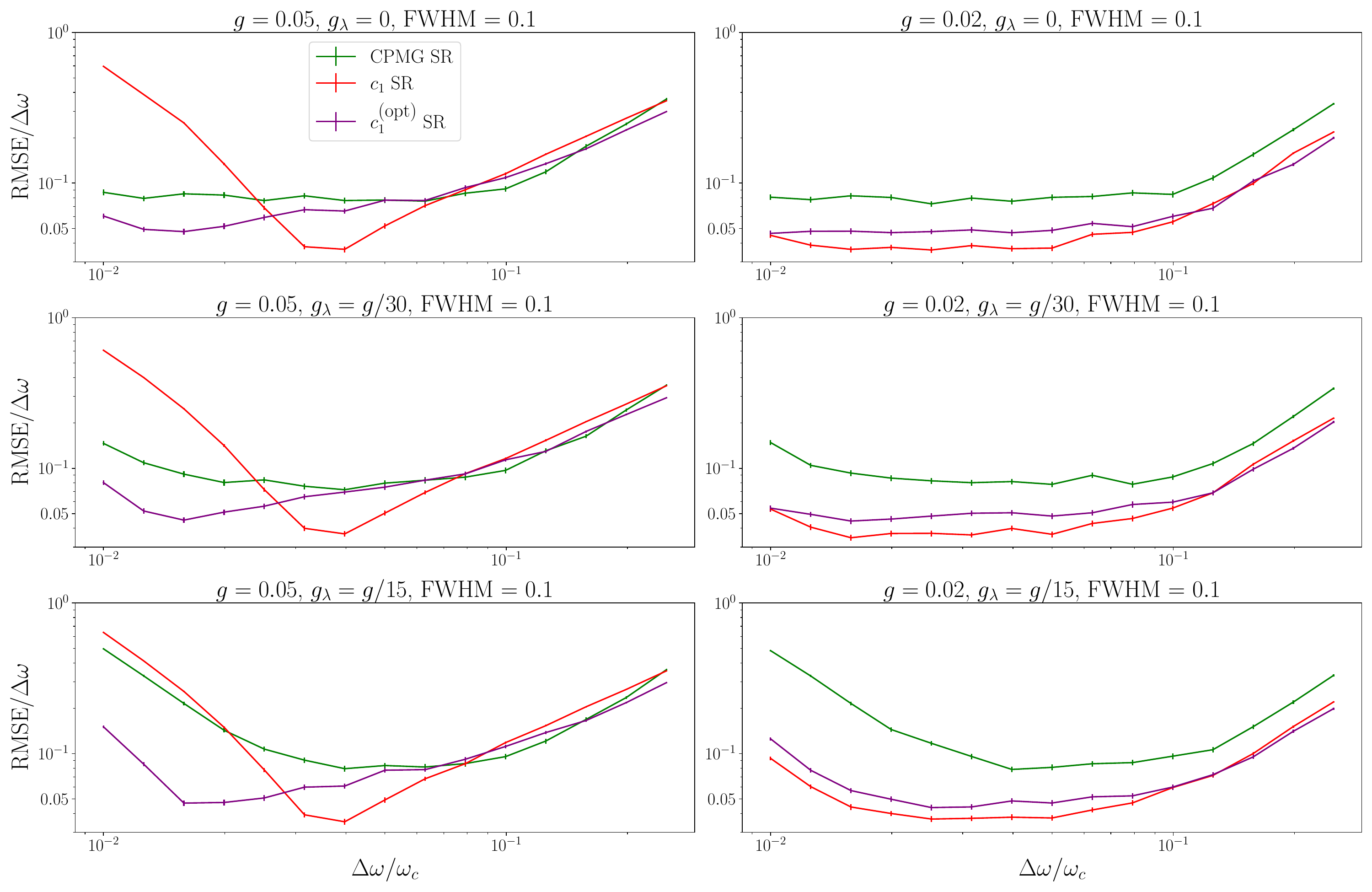}
    \caption{This plot is analogous to \cref{fig:MSE-numerics}, except we now also consider continuous control protocols.
        We numerically simulate the CPMG protocol, the \( c_1 \) protocol, and the optimized \( c_1^{\rm (opt)} \) protocol (where recall we optimized for a small and smooth control amplitude with small fourth filter function), all with \( \kappa =4 \). For each of the protocols, we utilize \( N_{\rm CPMG} \) samples for a desired noiseless relative error of \( \delta = 0.1 \) given in \cref{eq:number-measurements}.
        Note that all the protocols fail at large enough \( \Delta\omega \) because we are using a second order Taylor expansion approximation for \( \angles{P} \).
        We could go to higher order to remove this effect, but in this work we are primarily concerned with the limit where \( \Delta\omega \) is small.
        (Top left): Noiseless case and \( g=0.06 \). The Lorentzian noise parameters are set to \( g_\lambda = W = 0 \). As \( \Delta\omega\to 0 \), the instantaneous control protocols achieve the desired relative accuracy while the \( c_1 \) protocol fails due to the higher order filter functions. However, there is a range of \( \Delta\omega \) values where the \( c_1 \) protocol outperforms the instantaneous control protocols due to their filter functions having a larger second derivative at the centroid. Meanwhile, due to its small fourth filter function, the \( c_1^{\rm opt} \) protocol performs well for smaller values of \( \Delta\omega \).
        (Top right): Noiseless case and \( g=0.02 \). The Lorentzian noise parameters are set to \( g_\lambda = W = 0 \). This is similar to the top left case, except that now because \( g \) is smaller, the effect of the higher order filter functions is reduced, resulting in a larger range of values of \( \Delta\omega \) for which the continuous control protocols outperform the instantaneous control protocols.
        (Middle left): Noisy case (\( g_\lambda = g/30 \)) and \( g = 0.06 \). The Lorentzian noise parameters are set to \( g_\lambda = g/30 \) and FWHM \( W = 0.1 \).
        (Middle right): Noisy case (\( g_\lambda = g/30 \)) and \( g = 0.02 \). This case is the same as the Middle left, except again because of the smaller \( g \) value, the continuous control protocols perform better for longer.
        (Bottom left): Noisy case (\( g_\lambda = g/15 \)) and \( g = 0.06 \). The Lorentzian noise parameters are set to \( g_\lambda = g/15 \) and FWHM \( W = 0.1 \).
        (Bottom right): Noisy case (\( g_\lambda = g/15 \)) and \( g = 0.02 \). This case is the same as the bottom left, except again because of the smaller \( g \) value, the continuous control protocols perform better for longer.
    }
    \label{fig:cc-mse}
\end{figure*}

One major benefit of considering the superresolution criteria for continuous controls is that it easily allows for numerical optimization.
In particular, in this section, we will begin with \( c_1 \)-SR as an initial seed and optimize to find \( c_1^{\rm (opt)} \)-SR. We minimize the following Lagrangian,
\begin{widetext}
    \begin{equation}
        \calL = - F^{(2)\prime\prime}(\omega_c, \kappa\tau) + \mu_1 \int S_\lambda(\omega)F^{(2)}(\omega,\kappa\tau)\dd\omega  + \mu_2 F^{(2)}(\omega_c, \kappa\tau) + \mu_3 \norm{c(t)}_2^2 + \mu_4 F^{(4)}(\omega_c,\omega_c,\kappa\tau) + \mu_5 \norm{c'(t)}_2^2.
    \end{equation}
\end{widetext}
The first term represents maximizing the Fisher information (proxied by the second derivative of the filter function at the centroid).
The \( \mu_2 \) term enforces the superresolution condition that the filter function vanishes at the centroid -- in the case of instantaneous controls, this \emph{exactly} results in superresolution, whereas in the case of continuous controls, this results in superresolution up to order \( g^4 \).
The \( \mu_1 \) term means that we are attempting to minimize the overlap between the filter function and the Lorentzian noise, thus reducing the effect of the noise on our protocols.
The \( \mu_3 \) term encourages the resulting controls to have small amplitude.
The \( \mu_4 \) term tries to minimize the effect of the higher order filter functions.
Finally, the \( \mu_5 \) term enforces that the resulting controls are smooth. 
We discretize the time interval in increments of $\Delta t$, so that the control waveform $c(t)$ is parameterized as $\bm c = (c(0), c(\Delta t), c(2\Delta t), \dots)$. 
Using PyTorch tensors \cite{pytorch}, $\mathcal L$ can be quickly evaluated an optimized over as a function of $\bm c$.

We emphasize that the numerical optimization framework is flexible, and there are many different terms that one could add to the Lagrangian depending on the setting and task.
The Lagrangian we have chosen for illustration purposes attempts to find superresolution controls that are the most robust to noise, are small and smoothly varying, and have the largest regime of validity according to \cref{eq:cc-valid-regime}.
For example, if instead of Lorentzian noise, the relevant noise model is \( 1/f \) noise as is common in superconducting qubit platforms~\cite{paladino20141/f-noise:-impl}, then the form of \( \calL \) is the same except that \( S_\lambda \) now becomes the spectrum of \( 1/f \) noise.

We numerically optimize \( \calL \) with respect to $\bm c$ parameterizing the control \( c(t) \).
For illustration, we seed our optimizer with the \( c_1 \) protocol given in \cref{eq:c1-protocol}, though we replace the instantaneous pulse in the middle simply by a random small number. 
That is, the optimizer is seeded with $\bm c = (-1/2, -1/2, \dots, v, -1/2, -1/2, \dots)$, where $v$ is chosen randomly between $-1/2$ and $1/2$.
We will fix \( \kappa = 4 \).
After optimizing, we find new controls, which we call \( c_1^{\rm (opt)}(t) \).
The resulting control sequence is shown in \cref{fig:all-controls}, along with the control sequence for \( c_1 \) for comparison.
Also in \cref{fig:all-controls}, we show the corresponding filter functions for \( c_1^{\rm (opt)} \), \( c_1 \), and CPMG.
Note that one will get different results depending on what is being optimized for. In this example, to mimic realistic constraints on the amplitude of a control waveform as well as the it being more experimentally friendly to implement smoothly varying waveforms, we want \( \max_t \abs*{c_1^{\rm (opt)}(t)} \) to bounded and we want \( c_1^{\rm opt}(t) \) to be smooth.
We see that \( \max_t \abs*{c_1^{\rm (opt)}(t)}\) is bounded at \( \sim 3/4 \) and \( c_1^{\rm opt}(t) \) is smooth, whereas the ideal CPMG and \( c_1 \) protocols have infinite amplitude jumps. 
Despite restricting the maximum amplitude of the control waveform, the numerical optimization is still able to find a good control sequence that performs well.
Specifically, the \( c_1^{\rm (opt)}(t) \) and \( c_1 \) protocols both have a much larger value of \( F''(\omega_c, \kappa\tau) \) and have smaller overlap with low frequency Lorentzian noise, and \( c_1^{\rm (opt)}(t) \) is able to achieve this even with small amplitude controls.
Additionally, because we optimized \( c_1^{\rm opt} \) to have a small value of \( F^{(4)}(\omega_c, \omega_c, \kappa\tau) \), \( c_1^{\rm opt} \) will outperform \( c_1 \) at larger values of \( g \) where higher filter functions matter.

We simulate the protocols to produce the analogous plot as \cref{fig:MSE-numerics}, and all of these effects are shown in \cref{fig:cc-mse}.
We see that when \( \bigO{g^4} \) terms are irrelevant, both the \( c_1 \) and \( c_1^{\rm (opt)} \) protocol outperform CPMG, both in the noiseless and noisy cases.

\section{Entanglement enhancement}
\label{sec:entanglement}

In this section, and in \cref{ap:entanglement-advantage}, we show that entangled qubits and measurements can take any superresolution protocol that utilizes a single qubit and \( N \) repetitions and convert it to a superresolution protocol that achieves the same relative error \( \delta \) with \( N_e \sim \frac{\beta}{\Delta\omega^{1-\mu}} \) entangled qubits and
\( N_r \sim \frac{N}{N_e^2}\) repetitions, with $\beta>0$ and $\mu>0$ constants.
We argue that the relevant figure of merit is to compare \( N \) to \( N_e N_r \).
Therefore, we see that utilizing entanglement requires a factor of \( N/N_eN_r \sim \frac{\beta}{\Delta\omega^{1-\mu}} = N_e \) fewer resources.
Importantly, this comes with the caveat that the entangled qubits evolve under the Hamiltonian \( H_{N_e}(t) = \sum_{i=1}^{N_e} \parentheses{\gamma(t) \sigma_i^z + c(t) \sigma_i^x} \).
This means that the \( N_e \) qubits experience the \emph{same} realization of \( A_1,A_2,B_1, \) and \( B_2 \).

For simplicity, we will derive this entanglement advantage by utilizing GHZ states, but we expect it to work for other entangled states as well, such as Hamming weight \( N_e/2 \) Dicke states.
We therefore let the initial state of our $N_e$ qubit ensemble be \( \ket{\Psi_0} = \frac{1}{\sqrt 2}\parentheses{\ket0^{\otimes N_e} + \ket1^{\otimes N_e}} \), allow it to evolve under \( H_{N_e}(t) \) for a time \( \kappa\tau \), and measure the overlap of the final state with \( \ket{\Psi_0} \). 
Specifically, we measure the qubit in with the POVM $\{\ket{\Psi_0}\bra{\Psi_0}, \bbI - \ket{\Psi_0}\bra{\Psi_0} \}$, therefore producing a sample $\psi_i \in \{0,1\}$. 
Performing this $N_r$ times yields $(\psi_1,\dots,\psi_{N_r})$.
Thus, if \( U(t) \) is the time evolution operator under \( H(t) \), then \( U(t)^{\otimes N_e} \) is the time evolution operator under \( H_{N_e}(t) \).
We are therefore interested in the expected probability \( \angles{P_{N_e}} \) with
\begin{equation}
    \angles*{P_{N_e}} = \angles{\abs{\bra{\Psi_0}U(\kappa\tau)^{\otimes N_e}\ket{\Psi_0}}^2}.
\end{equation}
Restricting our attention to instantaneous controls, following the a similar approach as the \( N_e=1 \) calculation in \cref{ap:filter-function}, we find that
\( \angles{P_{N_e}} = \frac{1}{2} + \frac{1}{2}\e^{-N_e^2 \chi(\kappa\tau)}, \)
with \( \chi \) is the one qubit decay rate. 
Our estimator is therefore $\sqrt{\frac{1}{N_e^2 b} \parentheses*{1 - \frac{1}{N_r}\sum_{i=1}^{N_r} \psi_i}}$, which is identical to \cref{eq:estimator} with after $b \mapsto N_e^2 b$.
Computing the Fisher information then yields
\begin{equation}
    \lim_{\Delta\omega\to0}\FI_{\Delta\omega}(\angles*{P_{N_e}}) = N_e^2 \lim_{\Delta\omega\to0}\FI_{\Delta\omega}(\angles*{P}) = 4 N_e^2 b.
\end{equation}
In other words, the Fisher information for \( N_e \) entangled qubits is \( N_e^2 \) times that of a single qubit.
It follows from the Cram\'er Rao bound that in order to achieve a relative error of \( \varepsilon \), the protocol must repeated
\begin{equation}
    \label{eq:number-repetitions}
    N_r \sim \frac{1}{4 N_e^2 b \delta^2 \Delta\omega^2}
\end{equation}
times.

This result holds when $N_e$ is a constant independent of $\Delta\omega$, and gives a resource advantage of $\frac{N}{N_e, N_r} \sim N_e$. 
However, to push the entanglement advantage further, we can set $N_e$ to be a function of $\Delta\omega$ and redo the $\Delta\omega\to 0$ limit in the Fisher information.
Setting $N_e = \beta/\Delta\omega^{1-\mu}$ results in a number of repetitions
\begin{equation}
    N_r \sim \frac{1}{4b\beta^2\mu^2 \delta^2\Delta\omega^{2\mu}}.
\end{equation}
This gives an entanglement advantage of $\frac{N}{N_e N_r} \sim \mu^2 N_e = \frac{\mu^2 \beta}{\Delta\omega^{1-\mu}}$. 

Finally, the same result holds for continuous controls superresolution protocols in their regime of applicability -- that is, when the higher order filter functions can be neglected.

\section{Comparison to other methods}
\label{sec:comparisons}

In this section, we will compare superresolution protocols to other quantum and classical protocols.
In particular, in \cref{sec:qns}, we will describe standard quantum noise spectroscopy (QNS) techniques for estimating \( \Delta\omega \), and in \cref{sec:classical-methods}, we will perform a Fisher information analysis on classical superresolution protocols.

\subsection{Quantum noise spectroscopy}
\label{sec:qns}

In our superresolution protocols, we look for filter functions that vanish at the centroid \( \omega_c \) but have a large second derivative.
Traditional QNS methods, however, aim for filter functions with a large a narrow peak at a fixed frequency \( \omega^\ast \)~\cite{Szankowski2017Environmental-n}.
By choosing a set of control sequences that vary the location of $\omega^*$ along the $\omega$ axis, the overlap between the corresponding filter functions and the signal \( S(\omega) \) will vary, being largest when \( \omega^\ast \) is near \( \omega_1 \) or \( \omega_2 \).
In this way, one can estimate \( \Delta\omega = \omega_2-\omega_1 \).
In stark contrast to superresolution protocols, these QNS methods require \( \kappa \to \infty \) as \( \Delta\omega\to 0 \) in order for the peak in the filter function to be narrow than \( \sim\Delta\omega \) around \( \omega^\ast \).
Thus, superresolution protocols fundamentally outperform traditional QNS methods at estimating \( \Delta\omega \) as they only require the signal to be observed for a fixed, finite amount of time even as $\Delta\omega\to 0$.

\begin{figure}
    \centering
    \includegraphics[width=\linewidth]{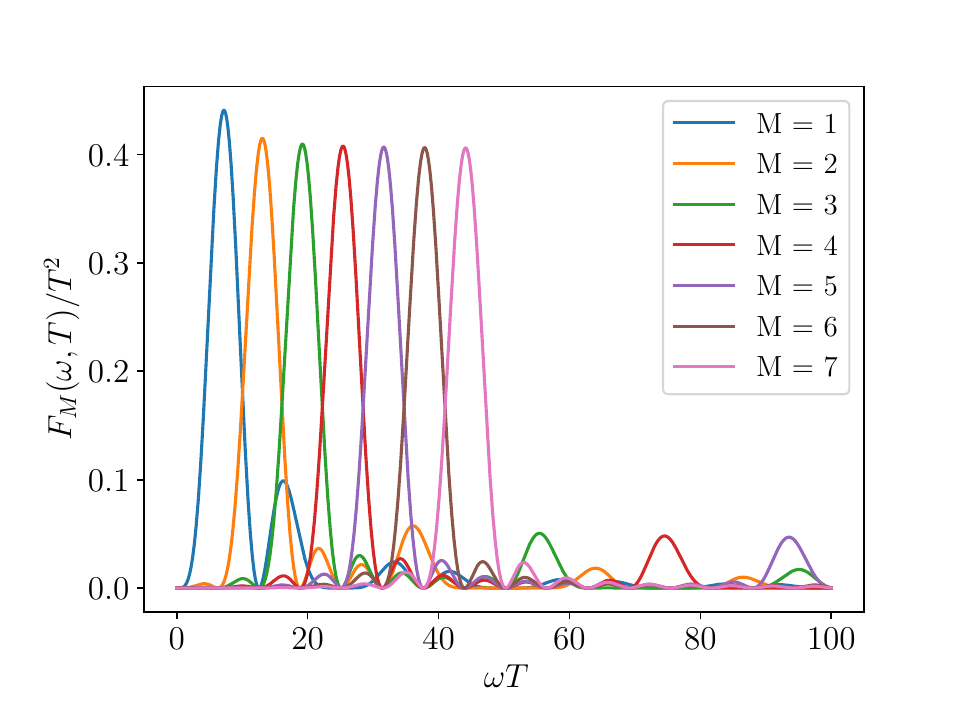}
    \caption{The filter functions given in \cref{eq:qns-filter-functions} for a sequence of \( M \) CPMG sequences in a total time \( T \). As \( M \) increases, the primary peaks slide along the frequency axis; as \( T \) increases the primary peaks become narrower in frequency.}
    \label{fig:qns-filter-function}
\end{figure}

To be more specific, we consider QNS using  CPMG sequences.
Given a total evolution time of \( T \), we consider the filter function arising from applying \( M \) CPMG sequences -- in this case, the controls consist of instantaneous \( \sigma^x \) gates at time \( T / (4M), 3T/(4M), 5T/(4M), \dots, (4M-1)T/(4M) \). 
The analytic form of the resulting filter function is
\begin{equation}
    \label{eq:qns-filter-functions}
    F_M(\omega, T) = \frac{16}{\omega^2} \sec^2\pargs{\frac{\omega T}{4M}}\sin^2\pargs{\frac{\omega T}{2}} \sin^4\pargs{\frac{\omega T}{8M}},
\end{equation}
and the resulting filter function is shown in \cref{fig:qns-filter-function}.

A key feature of \( F_M(\omega, T) \) is that it has a large peak at some \( \omega^\ast \) that depends on \( M \), and the width of this peak scales inversely with \( T \).
Thus, given the signal \( \gamma(t) \), we can evolve a qubit under \( \gamma(t) \) and the \( M \) CPMG sequences for a time \( T \) and measure the expectation value of \( \sigma^x \). This will yield the value of the overlap between the \( F_M(\omega, T) \) and \( S(\omega) \).
By performing this protocol for many different values of \( M \), we can estimate \( \omega_1 \) and \( \omega_2 \) based on when these expectation values are maximal.
However, the resulting resolution along the \( \omega \)-axis only scales as \( 1/T \) due to the standard Fourier uncertainty principle, so that \( T \) must scale at least as \( 1/\Delta\omega \) as \( \Delta\omega\to 0 \).

\subsection{Classical methods}
\label{sec:classical-methods}
Throughout this work, we have considered the signal \( \gamma(t) \) given in \cref{eq:signal}.
A superresolution protocol begins with the quantum state \( \ket+ \), evolves with the Hamiltonian in \cref{eq:hamiltonian} for a time \( \kappa\tau \), and then measures in the \( \sigma^x \) basis.
This protocol is repeated \( N \) times, where the \( \bm a = (A_1,A_2,B_1, B_2) \) in \( \gamma(t) \) are \emph{different} i.i.d.~ normally-distributed random variables for each of the \( N \) runs.
Such a procedure is applicable in a scenario where the signal is quasistatic over the timescale of a single shot, but varies from shot to shot, so that one is unable to make multiple measurements  with the \emph{same} \( \bm a \) coefficients.
In this section, we consider the case in which the signal is slowly varying, allowing one to make \( \calM \) measurements of \( \gamma(t) \) at fixed \( \bm a \).
This procedure is then repeated \( N/\calM \) times, resulting in \( N \) measurements of the signal.
We want to determine the best possible accuracy a classical algorithm can achieve in its estimate of \( \Delta\omega \) when given access to these \( N \) measurements. 

Before performing the analysis, we summarize the results.
When \( \calM = 1 \), the Fisher information vanishes, so that there is no algorithm that is able to achieve any accuracy in estimating \( \Delta\omega \). 
Note that our quantum superresolution protocols also operate in the \( \calM=1 \) regime and make \( N \) measurements; however, the key difference is that although the quantum procedures only measure once for a fixed realization of the \( \bm a \) coefficients just as the classical procedure, the quantum protocol can have nonzero Fisher information because the qubit is subjected to time evolution corresponding to the signal for \emph{all} times from \( 0 \) to \( \kappa\tau \) with a fixed \( \bm a \). 

On the other hand, we also consider the case when the coherence is long enough so that \( \calM > 1 \) measurements are able to be made with a fixed realization of \( \bm a \).
We show that if \( \calM \geq 4 \) and there is no noise (\( \lambda(t) = 0 \)), the classical Fisher information scales as \( \sim \frac{N}{\Delta\omega^2} \) as \( \Delta\omega\to 0 \), and hence a classical procedure is able to estimate \( \Delta\omega \) arbitrarily well as \( \Delta\omega\to 0 \).
We therefore see that the quantum superresolution protocols are best suited for when the coherence time is short (so that \( \calM \geq 4 \) is not practically achievable) and \( \Delta\omega \) is small.

\begin{figure}
    \centering
    \includegraphics[width=\linewidth]{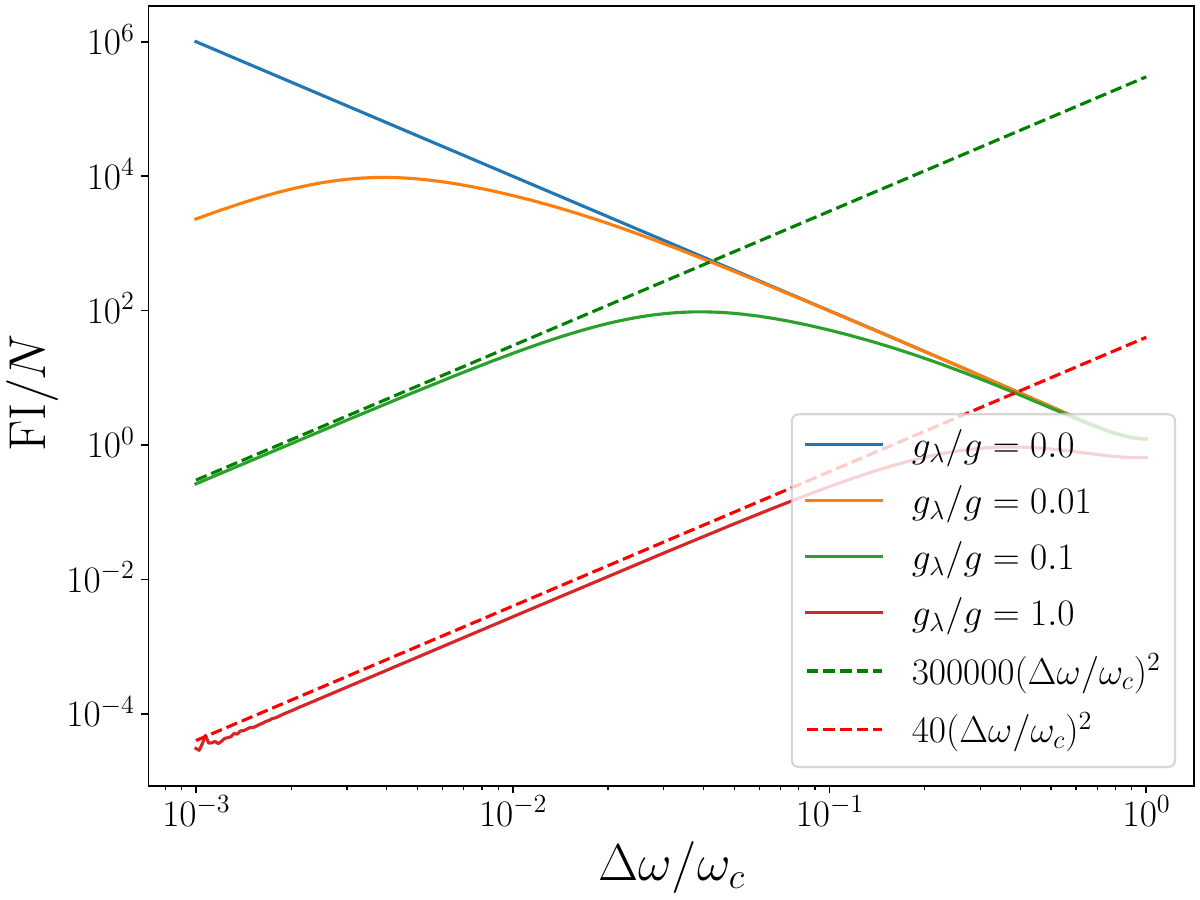}
    \caption{We plot \( \FI/N \) for the \( \calM=4 \) measurement block with Lorentzian noise with FWHM \( W=1/10 \) and strength \( g_\lambda \), \( \omega_c=1 \), and \( \kappa=1 \). For this plot, the measurement times were chosen to be \( t_m = (m-1)\kappa\tau/\calM \) for \( m=1,2,3,4 \).}
    \label{fig:classical-block-fi}
\end{figure}

Finally, we also consider the classical \( \calM\geq 4 \) case when subjected to Lorentzian noise. We find that the Fisher information scales as \( \sim \frac{N}{\Delta\omega^2} \) for large \( \Delta\omega \) and as \( \sim N \Delta\omega^2 \) as \( \Delta\omega\to 0 \), which is in fact the scaling that the quantum superresolution protocols achieve despite them only having access to a single measurement per realization of \( \bm a \).
The Fisher information as a function of \( \Delta\omega \) can be seen in \cref{fig:classical-block-fi} for various values of the noise strength.

\subsubsection*{Fisher information analysis}

In order to understand the performance of all possible classical methods, such as Fourier transform-based protocols or MUSIC \cite{hayes1996statistical},
we compute the classical Fisher information.
In the quantum protocols we have considered, a qubit is coupled to the signal and we have only measurement access through the qubit.
In contrast, for classical methods, we assume that we can directly measure the signal without disturbing it.
In other words, classical methods can take many samples from the signal, while quantum methods utilizes qubits that are continuously evolved by the signal but we only have one-time measure access to the qubit.

Because the Fisher information is additive, we can simply consider \( N = \calM \), and then multiply the end result by a factor of \( N/\calM \).
Given \( N = \calM \), we consider measuring the signal \( \gamma(t) \) at times \( 0 \leq t_1 < t_2 <  \dots < t_\calM \leq \kappa\tau \).
The result is a vector \( \bm s = (s_1,\dots, s_{\calM}) \) of measurement outcomes.
Specifically, from \cref{eq:signal}, in the noiseless case, we will have \( \bm s = g D \bm a \), where \( D \) is the \( \calM \times 4 \) matrix
\begin{equation}
    D = \begin{pmatrix}
        \cos\pargs{\omega_1 t_1} & \cos\pargs{\omega_2 t_1} &
        \sin\pargs{\omega_1 t_1} & \sin\pargs{\omega_2 t_1}                   \\
        \vdots                   & \vdots                   & \vdots & \vdots \\
        \cos\pargs{\omega_1 t_\calM} & \cos\pargs{\omega_2 t_\calM} &
        \sin\pargs{\omega_1 t_\calM} & \sin\pargs{\omega_2 t_\calM}
    \end{pmatrix}.
\end{equation}
In the presence of noise \( \lambda \), we instead have \( \bm s = g D \bm a + g_\lambda \bm \lambda \), where \( \bm\lambda = (\lambda_1,\dots,\lambda_\calM) \), and \( g_\lambda \lambda_m \) is the value of the noise at time \( t_m \).

In order the calculate the Fisher information, we are first interested in the probability density \( \rho(\bm s \vert \Delta\omega) \) that quantifies the probability of measuring a given \( \bm s \) over the random values of \( \bm a \) and the random values of \( \bm \lambda \).
The probability distribution over \( \bm a \) is simply a zero-mean, unit-variance independent Gaussian distribution on each factor, while the distribution over \( \bm \lambda \) is zero-mean multivariate Gaussian distribution with \( \calM \times \calM \) covariance matrix \( V^{-1} \) with entries
\( (V^{-1})_{ij} = \exp\bargs{-W \pi\abs{t_i-t_j}} \).
Thus, letting \( \delta^{(\calM)} \) denote the Dirac delta function on \( \calM \) elements, we have that
\begin{equation}
    \begin{aligned}
        \rho & (\bm s\vert  \Delta\omega) = \frac{\sqrt{\abs{\det V}}}{(2\pi)^2(2\pi)^{\calM/2}} \int_{\bbR^4} \dd\bm a \int_{\bbR^\calM} \dd \bm \lambda          \\
             & \times \exp\bargs{-\frac{1}{2}(\bm a^T \bm a + \bm \lambda^T V \bm \lambda)} \delta^{(\calM)}\pargs{g D \bm a + g_\lambda \bm \lambda - \bm s}.
    \end{aligned}
\end{equation}

We compute the Fisher information in two cases, \( \calM=1 \) and \( \calM = 4 \).
When \( \calM = 1 \), a straightforward Gaussian integral shows that \( \rho(\bm s\vert \Delta\omega) \) is independent of \( \Delta\omega \) even if \( g_\lambda = 0 \).
Therefore, the Fisher information is zero for all \( \Delta\omega \) and \( N \).
Intuitively, this is due to the randomness in the \( \bm a \) coefficients; measuring the signal \( N \) times with \( N \) different \( \bm a \) yields a sequence of random measurements with no information about the underlying signal.

Next, we consider the \( \calM = 4 \) case.
We assume that \( t_1,\dots,t_4 \) have been chosen so that \( D \) is an invertible \( 4\times 4 \) matrix.
For example, \( t_m = m\kappa\tau/4 \) yields an invertible \( D \) unless \( \kappa / \omega_c \) is an even integer.
In this case, the \( \delta^{(4)} \) distribution exactly cancels the integral over \( \bm a \), and the \( \bm \lambda \) integral can be performed as a standard multivariate Gaussian integral.
We then find that \( \rho(\bm s \vert \Delta\omega) \) is the probability density function of a multivariate zero-mean Gaussian with covariance matrix \( \Sigma \),
where \( \Sigma^{-1} = \sigma - g_\lambda^2 \sigma (V + g_\lambda^2 \sigma)^{-1}\sigma \) and \( \sigma = (D^{-1})^T D^{-1} \). 
Notice that $\Sigma$ is guaranteed to be invertible as long as $g_\lambda$ is small.

The Fisher information for a multivariate Gaussian distribution whose covariance matrix depends on an unknown parameter is well-known and takes the form~\cite[Eq.~(6)]{porat1986computation-of-}
\begin{equation}
    \FI = \frac{N}{8}\Tr\bargs{\Sigma^{-1}\Sigma'  \Sigma^{-1} \Sigma'},
\end{equation}
where \( \Sigma' \) denotes the entrywise derivative of \( \Sigma \) with respect to \( \Delta\omega \), and we put back the factor of \( N/\calM \).
Using that for any matrix \( A \), \( (A^{-1})' = - A^{-1}A' A^{-1} \),
we can substitute for \( \frac{\dd}{\dd \Delta\omega}(V + g_\lambda^2 \sigma)^{-1} \) and \( \frac{\dd}{\dd \Delta\omega} \sigma = \frac{\dd}{\dd \Delta\omega} (D^{-1 T} D^{-1}) \) in order to arrive at a numerically computable form of the Fisher information.
The result is shown in \cref{fig:classical-block-fi}.

In fact, in the noiseless case when \( g_\lambda = 0 \), the exact Fisher information can be computed using a computer symbolic algebra software. For example, when \( \kappa = 1 \), we find that
\begin{equation}
    \FI_{\kappa=1} = \frac{N \pi ^2}{8 \omega_c^2} \left(2 \csc ^2\pargs{\frac{\pi  \Delta \omega}{2 \omega_c}}-1\right) = \frac{N}{\Delta\omega^2} + \bigO{1}.
\end{equation}
To see why the Fisher information diverges as \( \Delta\omega\to 0 \) in the noiseless case, we consider a simple protocol when \( N = \calM = 5 \).
Note that of course the Fisher information can only increase with increasing \( \calM \), and therefore the \( \calM=5 \) Fisher information must also diverge as \( \Delta\omega\to 0 \).
To be concrete, we will let \( \kappa =1 \), and by assumption there is a single realization of \( \bm a \).
We measure the signal at times \( 0, \tau/5, 2\tau/5, 3\tau/5, 4\tau/5 \).
Then the signal that we measure will be \( \bm s = D \bm a  \).
This time, because \( \calM=5 \), \( D \) is \( 5\times 4 \);
As \( \Delta\omega\to 0 \), \( D =  D_0 + \Delta\omega  D_1 + \Delta\omega^2  D_2 + \bigO{\Delta\omega^3} \).
Hence, for small \( \Delta\omega \), after measuring the signal five times and getting \( \bm s \), we have a system of five equations for five unknowns (\( A_1,A_2,B_1,B_2 \) and \( \Delta\omega \)), \( D_0\bm a + \Delta\omega  D_1\bm a + \Delta\omega^2  D_2\bm a = \bm s \).
From here, we can simply solve for the unknowns in order to determine \( \Delta\omega \).
As \( \Delta\omega\to 0 \), the Taylor expansion of \( D \) becomes more and more acccurate, meaning our estimate of \( \Delta\omega \) will also become more and more accurate.
Therefore, in this limit, we can achieve arbitrarily good accuracy in our estimate with only five measurements.
Recall the Cram\'er Rao bound says that the optimal estimation error \( \varepsilon^2 \) behaves as \( \frac{1}{N \FI} \).
Because it is possible to achieve arbitrarily good accuracy (\emph{i.e.}~\( \varepsilon\to 0 \)) with \( N = \calM =5 \) as \( \Delta\omega\to 0 \), we see that the Fisher information must diverge in this limit.

In conclusion, we see that if the coherence time of the signal is long enough that multiple measurements of the signal are able to be made with a single realization of \( \bm a \), then simple classical methods will outperform quantum superresolution protocols.
On the other hand, in the short coherence time setting, quantum protocols are the better choice.

\section{Conclusion}
\label{sec:conclusion}

In this work, building off of Ref.~\cite{gefen2019overcoming-reso}, we studied quantum superresolution for frequency resolution and estimation.
We found explicit conditions on the filter function defined solely in terms of the control Hamiltonian for a protocol to exhibit superresolution.
Using these conditions, we find many such protocols, as well as show how to numerically optimize control sequences to find controls that exhibit superresolution and satisfy other conditions, such as robustness to a specified noise model, smoothness of controls, etc.
We further showed that any superresolution protocol can be improved with entanglement.
We compared quantum superresolution methods to standard QNS methods and other classical methods in order to understand the regime in which quantum superconducting protocols will outperform other methods.

The effects of various nonidealities---such as low frequency noise, errors in the centroid estimation, or errors in the duration of the protocol---on the accuracy of frequency superresolution protocols can be analyzed (see \cref{ap:error-analysis} and Ref.~\cite{gefen2019overcoming-reso}), with the upshot being that superresolution still holds as long as the strength of the nonidealities is \( \lesssim \Delta\omega \).
It is an important future direction to understand and characterize more nonidealities, such as finite width peaks and asymmetric frequency spectra.


\begin{acknowledgments}
    We thank Jake Bringewatt, Gregory Quiroz, and Thomas Ruekgauer for helpful discussions.
    The authors acknowledge support from the Internal Research and Development program of the Johns Hopkins Applied Physics Laboratory. 
    J.T.I. thanks the Joint Quantum Institute at UMD for support through a JQI fellowship.
\end{acknowledgments}
%


\appendix\onecolumngrid

\section{Filter function formalism}
\label{ap:filter-function}

\subsection{Review of instantaneous control filter functions}
\label{ap:instantaneous-filter-function}

In this section, we review filter functions~\cite{kofman2004unified-theory-,cywiifmmode-nelse-nfiski2008how-to-enhance-,paz-silva2014general-transfe,ball2015walsh-synthesiz} for instantaneous controls in order to derive \cref{eq:probability-inst}.
We are considering a qubit coupled via the Hamiltonian in \cref{eq:hamiltonian} when the controls \( c(t) \) are instantaneous.
Specifically, suppose that we initialize the qubit in the \( \ket + \coloneqq \frac{1}{\sqrt 2}(\ket 0 + \ket 1) \) state and subject it to the Hamiltonian \( H(t) = (\gamma(t) + \lambda(t)) \sigma^z  \) while applying instantaneous \( \sigma^x \) gates at times \( t_1 < \dots < t_M \). Define \( t_0 = 0 \).
For notational convenience, define \( \gamma(t) \coloneqq \gamma(t) + \lambda(t) \).
Then the state after a time \( t > t_M \) is
\begin{salign}
    \ket{\psi(t)}
    &= \e^{-\i \int_{t_M}^t \tilde\gamma(t') \sigma^z \dd t'} \prod_{j=M}^1 \parentheses{\sigma^x \e^{-\i \int_{t_{j-1}}^{t_j} \tilde\gamma(t') \sigma^z \dd t'}} \ket+ \\
    &= \e^{-\i \int_{t_M}^t \tilde\gamma(t') \sigma^z \dd t'} \sigma^x \e^{-\i \int_{t_{M-1}}^{t_M} \tilde\gamma(t') \sigma^z \dd t'}\sigma^x \e^{-\i \int_{t_{M-2}}^{t_{M-1}} \tilde\gamma(t') \sigma^z \dd t'}
    \prod_{j=M-2}^1 \parentheses{\sigma^x \e^{-\i \int_{t_{j-1}}^{t_j} \tilde\gamma(t') \sigma^z \dd t'}} \ket+ \\
    &= \e^{-\i \int_{t_M}^t \tilde\gamma(t') \sigma^z \dd t' + \i \int_{t_{M-1}}^{t_M} \tilde\gamma(t') \sigma^z \dd t' -\i \int_{t_{M-2}}^{t_{M-1}} \tilde\gamma(t') \sigma^z \dd t'}
    \prod_{j=M-2}^1 \parentheses{\sigma^x \e^{-\i \int_{t_{j-1}}^{t_j} \tilde\gamma(t') \sigma^z \dd t'}} \ket+ \\
    &= \dots \\
    &= \exp\bargs{-\i \int_{t_M}^t \tilde\gamma(t')\sigma^z \dd t' + \i \sum_{j=1}^M (-1)^{M-j} \int_{t_{j-1}}^{t_j} \tilde\gamma(t')\sigma^z \dd t' } \ket+,
\end{salign}
where in the last line, we used that if $M$ is odd, we can still sandwich that term with $\sigma^x$ since $\sigma^x \ket+ = \ket +$. Finally, we define the \emph{switching function} \( f(t) \) by
\begin{equation}
    f(t) = \begin{cases}
        1  & \text{if } t_M < t \text{ or } t_{j-1} < t < t_j \text{ for } M-j \text{ even }, \\
        -1 & \text{if } t_{j-1} < t < t_j \text{ for } M-j \text{ odd } .
    \end{cases}
\end{equation}
Then
\begin{equation}
    \label{eq:state}
    \ket{\psi(t)}
    =
    \exp\bargs{-\i \int_0^t f(t')\tilde\gamma(t')\sigma^z \dd t'} \ket+ .
\end{equation}
The probability \( P(t) \coloneqq \abs{\bra{+}\ket{\psi(t)}}^2 \) of measuring \( 1 \) in the \( \sigma^x \) basis at time \( t \) is then
\begin{equation}
    P(t) = \cos^2\pargs{\int_0^t f(t')\tilde\gamma(t') \dd t'}.
\end{equation}

We recall that \( \gamma(t) \) is a random variable. The expectation \( \angles{P(t)} \) of \( P(t) \) is
\begin{salign}[eq:avg-P+]
    \angles{P(t)}
    &= \frac{1}{2} + \frac{1}{4}\parentheses{\angles{\e^{2\i \int_0^t f(t')\tilde\gamma(t') \dd t'}} + \angles{\e^{-2\i \int_0^t f(t')\tilde\gamma(t') \dd t'}}} \\
    &= \frac{1}{2} + \frac{1}{2} \e^{- 2 \int_0^t\int_0^t f(t')f(t'')\angles{\tilde\gamma(t')\tilde\gamma(t'')} \dd t' \dd t''} \\
    &= \frac{1}{2} + \frac{1}{2} \e^{- 2 \int_0^t\int_0^t f(t')f(t'') (C(t',t'') + C_\lambda(t', t'')) \dd t' \dd t''} \\
    &= \frac{1}{2} + \frac{1}{2} \e^{- 2 \int_0^t\int_0^t f(t')f(t'') (C(0,t''-t') + C_\lambda(0,t''-t')) \dd t' \dd t''} \\
    &= \frac{1}{2} + \frac{1}{2} \e^{-\frac{1}{\pi} \int_{-\infty}^\infty \dd\omega (S(\omega)+S_\lambda(\omega)) \int_0^t\int_0^t f(t')f(t'') \e^{\i\omega(t''-t')} \dd t' \dd t''} \\
    &= \frac{1}{2} + \frac{1}{2} \e^{-\frac{1}{\pi} \int_{-\infty}^\infty (S(\omega) + S_\lambda(\omega))F(\omega, t) \dd\omega},
\end{salign}
where we defined the \( C \) and \( C_\lambda \) to be the correlations for the Gaussian processes \( \gamma \) and \( \lambda \), used that \( \angles{\gamma(t)} = 0 \) and that for any Gaussian random variable \( X \), \( \angles{\e^{\alpha X}} = \e^{\alpha \angles X + \alpha^2 \angles{X^2} / 2} \).
Finally, we defined the \emph{filter function}~\cite{kofman2004unified-theory-,cywiifmmode-nelse-nfiski2008how-to-enhance-,paz-silva2014general-transfe,ball2015walsh-synthesiz}
\begin{equation}
    \label{eq:filter-function-switching-function}
    F(\omega, t) = \abs{\int_{0}^t f(t') \e^{\i\omega t'} \dd t'}^2 .
\end{equation}
We define \( \chi(t) \coloneqq \frac{1}{\pi}\int S(\omega)F(\omega, t)\dd\omega \), and similarly define \( \chi_\lambda(t) \), so that \( \angles{P(t)} = \frac{1}{2}+\frac{1}{2}\e^{-\chi(t) - \chi_\lambda(t)} \), thus reproducing \cref{eq:probability-inst} as desired.
The key point of the filter function formalism is that the details of the signal \( \gamma \) are entirely encoded by \( S(\omega) \), the noise \( \lambda \) by \( S_\lambda(\omega) \), and the details of our control sequence are entirely encoded by the filter function \( F(\omega, t) \).

\( \angles{P(t)} \) encodes the parameter \( \Delta\omega \). Then we can determine the Fisher information \( \FI_{\Delta\omega}(\angles{P(t)}) \) of \( \angles{P(t)} \) with respect to \( \Delta\omega \). The Fisher information tells us how accessible the information about the parameter is given access to the distribution. The Fisher information in the noiseless case is
\begin{salign}[eq:fisher-information-def]
    \FI_{\Delta\omega}(\angles{P(t)})
    &= \frac{1}{\angles{P(t)}} \parentheses{\frac{\partial \angles{P(t)}}{\partial \Delta\omega}}^2 + \frac{1}{1-\angles{P(t)}} \parentheses{\frac{\partial (1-\angles{P(t)})}{\partial \Delta\omega}}^2 \\
    &= \parentheses{\frac{\partial \angles{P(t)}}{\partial \Delta\omega}}^2
    \frac{1}{\angles{P(t)}(1-\angles{P(t)})} \\
    &= \frac{1}{2}\parentheses{\coth\chi(t)-1}\parentheses{\frac{\partial \chi(t)}{\partial \Delta\omega} }^2 .
\end{salign}

Given the spectrum given in \cref{eq:signal-S-omega} for the signal \( \gamma(t) \),
\begin{equation}
    \chi(t) = g^2 \parentheses{F(\omega_1, t) + F(\omega_2, t) + F(-\omega_1, t) + F(-\omega_2, t)} .
\end{equation}
With this signal, we are interested in the parameter \( \Delta\omega = \omega_2 - \omega_1 \).
Assuming \( \Delta\omega \) is small, this can be expanded around the centroid \( \omega_c = (\omega_1 + \omega_2)/2 \) as
\begin{equation}
    \chi(t) = g^2 \parentheses{ 4 F(\omega_c, t) + \frac{1}{2}\Delta\omega^2 F''(\omega_c, t)} + \bigO{g^2 \Delta\omega^4},
\end{equation}
where we used that \( F(\omega, t) = F(-\omega, t) \), and we denote by \( F'' \) the second derivative of \( F \) with respect to \( \omega \). It follows that
\begin{equation}
    \FI_{\Delta\omega}(\angles{P(t)})
    = \begin{cases}
        g^2 F''(\omega_c, t) + \bigO{\Delta\omega^2}                                                                               & \text{if } F(\omega_c, t) = 0    \\
        \frac{1}{2} g^4 F''(\omega_c, t)^2 \brackets{ \coth\pargs{4g^2 F(\omega_c, t)} - 1 }\Delta\omega^2 + \bigO{\Delta\omega^4} & \text{if } F(\omega_c, t) \neq 0
    \end{cases} .
\end{equation}

Recall that we formally defined a procedure to exhibit \emph{superresolution} if \( \lim_{\Delta\omega\to 0} \FI_{\Delta\omega}(\angles{P(t)}) > 0 \). We have therefore found a characterization of superresolution in terms of only the controls (\emph{i.e.}~the filter function). A sequence of control \( \sigma^x \) gates characterized by the switching function \( f(t) \) gives rise to superresolution if and only if the filter function vanishes as the centroid, \( F(\omega_c, t) = 0 \).
In summary, our goal in designing a superresolution protocol is to design a control sequence such that \( F(\omega_c, t) = 0 \) and \( F''(\omega_c, t) \) is maximized.

We immediately find such superresolution protocols by consider free and CPMG evolution.
Define the characteristic timescale \( \tau\coloneqq 2\pi/\omega_c \). In the \emph{free evolution} protocol, we apply no controls, so that the switching function \( f_{\rm free}(t) = 1 \) for all \( t \). From \cref{eq:filter-function-switching-function}, the filter function is therefore
\begin{equation}
    F_{\rm free}(\omega, \kappa \tau) = \frac{4}{\omega^2}\sin^2\pargs{\frac{\pi \kappa \omega}{\omega_c} } .
\end{equation}
In the \emph{CPMG protocol}, we assume that \( \kappa \) is an even integer, and our control is a sequence of \( \kappa/2 \) CPMG sequences; that is, we apply \( \sigma^x \) gates at times \( \tau/2,3\tau/2, 5\tau/2,\dots, (2\kappa-1)\tau / 2 \). The switching function is thus \( f_{\rm CPMG}(t) = \operatorname{sgn} \cos (\pi t / \tau) \), where \( \operatorname{sgn} \) is the sign function. From \cref{eq:filter-function-switching-function}, the filter function is
\begin{salign}
    F_{\rm CPMG}(\omega, \kappa \tau)
    &= \frac{1}{\omega^2}\abs{(1 - \e^{\i\omega \tau/2}) + (\e^{\i \omega (2\kappa-1)\tau/2} - \e^{\i\omega \kappa \tau}) + \sum_{j=1}^{\kappa - 1} (-1)^j (\e^{\i \omega (2j-1)\tau/2} - \e^{\i \omega (2j+1)\tau/2} ) }^2 \\
    &= \frac{16}{\omega^2} \sin ^4\!\left(\frac{\pi  \omega }{2 \omega_c}\right) \sec ^2\!\left(\frac{\pi  \omega }{\omega_c}\right) \sin ^2\!\left(\frac{\pi  \kappa  \omega }{\omega_c}\right) \\
    &= 4 F_{\rm free}(\omega, \kappa \tau) \sin^4\pargs{\frac{\pi\omega}{2\omega_c}} \sec^2\pargs{\frac{\pi\omega}{\omega_c}} .
\end{salign}
We therefore see that the free evolution protocol exhibits superresolution when \( \kappa \in \bbZ \) and the CPMG protocol exhibits superresolution when \( \kappa \in 2\bbZ \) because \( F_\ast(\omega_c, \kappa \tau) = 0 \). We then find that the Fisher informations are exactly as in \cref{eq:free-cpmg-fi}.


\subsection{Review of continuous control filter functions}
\label{ap:continuous-filter-function}

In \cref{ap:instantaneous-filter-function}, we reviewed the filter function formalism for a qubit coupled to a time-dependent signal and instantaneous \( \sigma^x \) \( \pi \)-pulse controls.
Using the Magnus expansion from Ref.~\cite{green2013arbitrary-quant}, we will now analyze the case when controls are allowed to be more general.
For simplicity, we will consider the noiseless case \( \lambda(t) = 0 \) in this derivation.
We discuss the addition of noise at the end.

As before, our signal plus noise is coupled to a qubit via \( H_0(t) = \gamma(t) \sigma^z \). We suppose furthermore that we supply a control Hamiltonian \( H_c(t) = c(t) \sigma^x \), so that the full Hamiltonian is \( H(t) = H_0(t) + H_c(t) \), as in \cref{eq:hamiltonian}.
In the case of instantaneous controls as above, \( c(t) \) is simply a sum of delta functions.
Our goal is, as before, to find \( P(t) \coloneqq \abs{\bra + U(t) \ket +}^2 \), where \( U(t) = \calT\!\exp\bargs{-\int_0^t H(t') \dd t'} \), with \( \calT\!\exp \) denoting the time ordered exponential.
We introduce the control propagator
\begin{equation}
    U_c(t) = \exp\pargs{-\int_0^t H_c(t')\dd t'} = \cos(\theta_c(t))\bbI - \i \sin(\theta_c(t)) \sigma^x, \qquad \theta_c(t) = \int_0^t c(t') \dd t',
\end{equation}
meaning that \( U(t) = U_c(t) \tilde U(t) \), where
\begin{equation}
    \tilde U(t) = \calT\!\exp\bargs{- \int_0^t \tilde H(t') \dd t'}, \qquad \tilde H(t) = U_c(t)^\dag H_0(t) U_c(t) = \gamma(t) \parentheses{
        \cos(2\theta_c(t))\sigma^z +
        \sin(2\theta_c(t)) \sigma^y
    }.
\end{equation}
Define the vector \( \bm a(t) = a(t) \hat{\bm a}(t) \), where \( \hat{\bm a} \) is a unit vector, such that \( \tilde U(t) = \e^{-\i a(t) \hat{\bm a}(t) \cdot \bm \sigma} \).
Then we have
\begin{salign}
    P(t)
    &= \abs*{\bra + U_c(t)\tilde U(t) \ket +}^2  \\
    &= \abs*{\bra + \tilde U(t) \ket +}^2 \\
    &= \abs*{ \cos a(t) - \i \sin a(t) \hat{\bm a}(t) \cdot \bra + \bm \sigma \ket + }^2 \\
    &= \abs*{ \cos a(t) - \i \sin a(t) a_x(t) }^2 \\
    &= \cos^2 a(t) + a_x(t)^2 \sin^2 a(t) ,
\end{salign}
where we defined \( \hat{\bm a}(t) = (a_x(t), a_y(t), a_z(t)) \).

Thus, the remaining task is to calculate \( \bm a(t) \).
We use the results from Ref.~\cite{green2013arbitrary-quant} to compute \( a(t) \).
With a first order Magnus expansion, we find that
\begin{equation}
    \bm a^{(1)}(t) = \int_0^t \gamma(t') \parentheses{0, \sin(2\theta_c(t')), \cos(2\theta_c(t'))} \dd t'.
\end{equation}
The second order term in the Magnus expansion yields
\begin{equation}
    \bm a^{(2)}(t) = \int_0^t \dd t_1\int_0^{t_1} \dd t_2 ~\gamma(t_1)\gamma(t_2) \parentheses{\sin\pargs{2\theta_c(t_1) - 2\theta_c(t_2)}, 0, 0}.
\end{equation}

Thus, up to order \( g^4 \), we have that \footnote{Note that odd order terms have vanished because \( \gamma \) has zero mean, and we ignore all terms of order \( g^6 \) or smaller.}
\begin{salign}
    \angles*{P(t)}
    &= 1 - \angles*{a(t)^2} + \frac{1}{3}\angles*{a(t)^4} + \angles*{a_x(t)^2 a(t)^2}
    = 1- \angles*{\norm*{\bm a^{(1)}}^2} - \angles*{\norm*{\bm a^{(2)}}^2} + \frac{1}{3}\angles*{\norm*{\bm a^{(1)}}^4} \\
    \begin{split}
         & = 1 - \int_0^t \dd t_1 \int_0^t \dd t_2 \angles*{\gamma(t_1)\gamma(t_2)}  \cos(2\theta_c(t_1) - 2 \theta_c(t_2))                             \\
         & \qquad -
        \int_0^t \dd t_1\int_0^{t_1} \dd t_2 \int_0^t \dd t_3\int_0^{t_3} \dd t_4
        ~\angles*{\gamma(t_1)\gamma(t_2) \gamma(t_3)\gamma(t_4)} \sin\pargs{2\theta_c(t_1) - 2\theta_c(t_2)}\sin\pargs{2\theta_c(t_3) - 2\theta_c(t_4)} \\
         & \quad + \frac{1}{3}\int_0^t \dd t_1 \int_0^t \dd t_2\int_0^t \dd t_3 \int_0^t \dd t_4 \angles*{\gamma(t_1)\gamma(t_2)\gamma(t_3)\gamma(t_4)}
        \cos(2\theta_c(t_1) - 2 \theta_c(t_2))
        \cos(2\theta_c(t_3) - 2 \theta_c(t_4))
    \end{split} \\
    \begin{split}
         & = 1 - \int_0^t \dd t_1 \int_0^t \dd t_2
        C(0, t_2-t_1)
        \cos(2\theta_c(t_1) - 2 \theta_c(t_2))                                                   \\
         & \quad -
        \int_0^t \dd t_1\int_0^{t_1} \dd t_2 \int_0^t \dd t_3\int_0^{t_3} \dd t_4                \\
         & \qquad\qquad \parentheses{
            C(0, t_2 - t_1)C(0, t_4 - t_3)
            + C(0, t_3 - t_1)C(0, t_4 - t_2)
            + C(0, t_4 - t_1)C(0, t_3 - t_2)
        }                                                                                        \\
         & \qquad \qquad
        \sin\pargs{2\theta_c(t_1) - 2\theta_c(t_2)}\sin\pargs{2\theta_c(t_3) - 2\theta_c(t_4)}   \\
         & \quad + \frac{1}{3}\int_0^t \dd t_1 \int_0^t \dd t_2\int_0^t \dd t_3 \int_0^t \dd t_4 \\
         & \qquad\qquad
        \parentheses{
            C(0, t_2 - t_1)C(0, t_4 - t_3)
            + C(0, t_3 - t_1)C(0, t_4 - t_2)
            + C(0, t_4 - t_1)C(0, t_3 - t_2)
        }                                                                                        \\
         & \qquad\qquad
        \cos(2\theta_c(t_1) - 2 \theta_c(t_2))
        \cos(2\theta_c(t_3) - 2 \theta_c(t_4))
    \end{split} \\
    \begin{split}
         & = 1 - \frac{1}{2\pi}\int \dd\omega S(\omega) \int_0^t \dd t_1 \int_0^t \dd t_2 \e^{\i\omega(t_2-t_1)}
        \cos(2\theta_c(t_1) - 2 \theta_c(t_2))                                                                   \\
         & \quad -
        \int_0^t \dd t_1\int_0^{t_1} \dd t_2 \int_0^t \dd t_3\int_0^{t_3} \dd t_4                                \\
         & \qquad\qquad \parentheses{
            C(0, t_2 - t_1)C(0, t_4 - t_3)
            + C(0, t_3 - t_1)C(0, t_4 - t_2)
            + C(0, t_4 - t_1)C(0, t_3 - t_2)
        }                                                                                                        \\
         & \qquad \qquad
        \sin\pargs{2\theta_c(t_1) - 2\theta_c(t_2)}\sin\pargs{2\theta_c(t_3) - 2\theta_c(t_4)}                   \\
         & \quad + \frac{1}{3}\int_0^t \dd t_1 \int_0^t \dd t_2\int_0^t \dd t_3 \int_0^t \dd t_4
        C(0, t_2 - t_1)C(0, t_4 - t_3)
        \Bigg[                                                                                                   \\
         & \qquad\qquad
            \cos(2\theta_c(t_1) - 2 \theta_c(t_2))\cos(2\theta_c(t_3) - 2 \theta_c(t_4))
        + \cos(2\theta_c(t_1) - 2 \theta_c(t_3))\cos(2\theta_c(t_2) - 2 \theta_c(t_4))                           \\
         & \qquad\qquad + \cos(2\theta_c(t_1) - 2 \theta_c(t_4))\cos(2\theta_c(t_2) - 2 \theta_c(t_3))
            \Bigg]
    \end{split} \\
    &= 1 - \frac{1}{2\pi}\int \dd\omega S(\omega) F^{(2)}(\omega, t)
    - \frac{1}{12\pi^2}\int \dd\omega_1 \int \dd\omega_2 S(\omega_1) S(\omega_2)
    F^{(4)}(\omega_1, \omega_2, t),
    \label{eq:probability-continuous-controls}
\end{salign}
where we defined
\begin{salign}
    F^{(2)}(\omega, t)
    &= \int_0^t \dd t_1 \int_0^t \dd t_2 \e^{\i\omega(t_2-t_1)}
    \cos(2\theta_c(t_1) - 2 \theta_c(t_2)) \\
    %
    %
    &= \abs{\int_0^t \cos\pargs{2\theta_c(t')} \e^{\i \omega t'} \dd t'}^2
    + \abs{\int_0^t \sin\pargs{2\theta_c(t')} \e^{\i \omega t'} \dd t'}^2
\end{salign}
and
\begin{equation}
    \label{eq:fourth-filter-function}
    \begin{split}
        F^{(4)}(\omega_1, \omega_2, t)
         & =
        -\int_0^t \dd t_1 \int_0^t \dd t_2\int_0^t \dd t_3 \int_0^t \dd t_4
        \e^{\i\omega_1(t_2-t_1)}\e^{\i\omega_2(t_4-t_3)}
        \Bigg[                                                                                         \\
         & \qquad\qquad
            \cos(2\theta_c(t_1) - 2 \theta_c(t_2))\cos(2\theta_c(t_3) - 2 \theta_c(t_4))
        + \cos(2\theta_c(t_1) - 2 \theta_c(t_3))\cos(2\theta_c(t_2) - 2 \theta_c(t_4))                 \\
         & \qquad\qquad + \cos(2\theta_c(t_1) - 2 \theta_c(t_4))\cos(2\theta_c(t_2) - 2 \theta_c(t_3))
        \Bigg]                                                                                         \\
         & +3
        \int_0^t \dd t_1 \int_0^{t_1} \dd t_2\int_0^t \dd t_3 \int_0^{t_3} \dd t_4
        \e^{\i\omega_1(t_2-t_1)}\e^{\i\omega_2(t_4-t_3)}
        \Bigg[                                                                                         \\
         & \qquad\qquad
            \sin(2\theta_c(t_1) - 2 \theta_c(t_2))\sin(2\theta_c(t_3) - 2 \theta_c(t_4))
        + \sin(2\theta_c(t_1) - 2 \theta_c(t_3))\sin(2\theta_c(t_2) - 2 \theta_c(t_4))                 \\
         & \qquad\qquad + \sin(2\theta_c(t_1) - 2 \theta_c(t_4))\sin(2\theta_c(t_2) - 2 \theta_c(t_3))
            \Bigg].
    \end{split}
\end{equation}
The upshot is then \cref{eq:probability-continuous-controls}, which Taylor expands to \cref{eq:continuous-probability}.

We can now see that this indeed reproduces the instantaneous case from above; when the controls are instantaneous, \( c(t) \) takes the form \( c(t) = \frac{\pi}{2} \sum_{i=1}^M \delta(t - t_i)  \). In this case, \( \theta_c(t) = \frac{\pi}{2} (\text{num of pulses before time }t) \), so that \( \sin\pargs{2\theta_c(t')} = 0 \) and
\( \cos\pargs{2\theta_c(t')} = (-1)^{\text{num of pulses before time }t} \). Note that \( \cos\pargs{2\theta_c(t')} \) is exactly the switching function \( f(t) \).
We see that \( F^{(2)}(\omega, t) \) exactly matches \cref{eq:filter-function-switching-function}, as expected. Similarly, \( F^{(4)}(\omega_1, \omega_2, t) = 3F^{(2)}(\omega_1, t) F^{(2)}(\omega_2, t) \).

Finally, we consider adding a nonzero noise \( \lambda(t) \neq 0 \).
Exactly as in the instantaneous control case, the lowest order addition to \cref{eq:probability-continuous-controls} is simply \( \frac{1}{2\pi} \int \dd\omega S_\lambda(\omega)F^{(2)}(\omega, t) \).
At higher order, we have
\( \frac{-1}{12\pi^2}\int \dd\omega_1 \int \dd\omega_2 S_\lambda(\omega_1) S_\lambda(\omega_2)
F^{(4)}(\omega_1, \omega_2, t) \)
along with the cross term
\( \frac{-1}{12\pi^2}\int \dd\omega_1 \int \dd\omega_2 S_\lambda(\omega_1) S(\omega_2)
F^{(4)}(\omega_1, \omega_2, t) \).


\subsection{Local optimality of instantaneous controls for superresolution}

In this short section, we show that optimizing the Fisher information by tuning the filter functions results in instantaneous controls being \emph{local} optima.
In particular, we would like to minimize the Lagrangian
\begin{salign}
    \calL
    &= -F^{(2)\prime\prime}(\omega_c, \kappa \tau) + \lambda F^{(2)}(\omega_c, \kappa\tau) \\
    &= \int_0^{\kappa \tau}\dd t_1 \int_0^{\kappa \tau} \dd t_2 \parentheses{(t_2-t_1)^2 + \lambda} \cos(\omega_c (t_2-t_1))\cos\pargs{2\theta_c(t_2) - 2 \theta_c(t_1)} ,
\end{salign}
where we are assuming that we are in the regime where \( F^{(4)} \) can be ignored, and \( \lambda \) is a Lagrange multiplier enforcing that the filter function vanishes at the centroid \( \omega=\omega_c \).
Because \( \frac{\dd}{\dd t}\theta_c(t) = c(t) \), we can consider minimizing \( \calL \) with respect to \( \theta_c(t) \).
Performing a variation \( \delta \theta_c(t) \), we find
\begin{salign}
    \begin{split}
        \delta \calL
         & = \int_0^{\kappa \tau}\dd t_1 \int_0^{\kappa \tau} \dd t_2 \parentheses{(t_2-t_1)^2 + \lambda} \cos(\omega_c (t_2-t_1)) \\
         & \qquad\brackets{
            \cos\pargs{2\theta_c(t_2) - 2 \theta_c(t_1) + 2 \delta\theta_c(t_2) - 2 \delta\theta_c(t_1) }
            -\cos\pargs{2\theta_c(t_2) - 2 \theta_c(t_1) }
        }
    \end{split} \\
    &= -2\int_0^{\kappa \tau}\dd t_1 \int_0^{\kappa \tau} \dd t_2 \parentheses{(t_2-t_1)^2 + \lambda} \cos(\omega_c (t_2-t_1))
    \sin\pargs{2\theta_c(t_2) - 2 \theta_c(t_1) } \parentheses{\delta\theta_c(t_2) - \delta\theta_c(t_1)} \\
    \begin{split}
         & = -2\int_0^{\kappa \tau}\dd t_1 \int_0^{\kappa \tau} \dd t_2 \parentheses{(t_2-t_1)^2 + \lambda} \cos(\omega_c (t_2-t_1))
        \sin\pargs{2\theta_c(t_2) - 2 \theta_c(t_1) } \delta\theta_c(t_2)                                                                 \\
         & \qquad +2\int_0^{\kappa \tau}\dd t_1 \int_0^{\kappa \tau} \dd t_2 \parentheses{(t_2-t_1)^2 + \lambda} \cos(\omega_c (t_2-t_1))
        \sin\pargs{2\theta_c(t_2) - 2 \theta_c(t_1) } \delta\theta_c(t_1)
    \end{split}\\
    \begin{split}
         & = -2\int_0^{\kappa \tau}\dd t_1 \int_0^{\kappa \tau} \dd t_2 \parentheses{(t_2-t_1)^2 + \lambda} \cos(\omega_c (t_2-t_1))
        \sin\pargs{2\theta_c(t_2) - 2 \theta_c(t_1) } \delta\theta_c(t_2)                                                                 \\
         & \qquad +2\int_0^{\kappa \tau}\dd t_1 \int_0^{\kappa \tau} \dd t_2 \parentheses{(t_2-t_1)^2 + \lambda} \cos(\omega_c (t_2-t_1))
        \sin\pargs{2\theta_c(t_1) - 2 \theta_c(t_2) } \delta\theta_c(t_2)
    \end{split}\\
    &= 4\int_0^{\kappa \tau}\dd t_1 \int_0^{\kappa \tau} \dd t_2 \parentheses{(t_2-t_1)^2 + \lambda} \cos(\omega_c (t_2-t_1))
    \sin\pargs{2\theta_c(t_1) - 2 \theta_c(t_2) } \delta\theta_c(t_2) \\
    &= 4 \int_0^{\kappa \tau} \dd t_2 \brackets{
        \int_0^{\kappa \tau}\dd t_1
        \parentheses{(t_2-t_1)^2 + \lambda} \cos(\omega_c (t_2-t_1))
        \sin\pargs{2\theta_c(t_1) - 2 \theta_c(t_2) }
    } \delta\theta_c(t_2) .
\end{salign}
Therefore, in order for the controls \( c(t) \) to be a local optimum, we need
\( F(\omega_c,\kappa\tau) = 0 \) and
\begin{equation}
    \forall t\colon ~ \int_0^{\kappa \tau}\dd t_1
    \parentheses{(t-t_1)^2 + \lambda} \cos(\omega_c (t-t_1))
    \sin\pargs{2\theta_c(t_1) - 2 \theta_c(t) } = 0 .
\end{equation}
The latter condition is \emph{always} satisfied for instantaneous controls, because for instantaneous controls, \( 2\theta_c(t) \in \pi \bbZ \) for all \( t \).

\section{Robustness of the superresolution criteria}
\label{ap:robustness}

In this section, we show that there is no measurement basis or initial state such that superresolution can be achieved without \( F(\omega_c,\kappa\tau) = 0 \). In other words, the superresolution criteria that the filter function vanishes at the centroid is robust.

Suppose we have an instantaneous protocol defined by the switching function \( f(t) \), and suppose that the noise is \( \lambda(t) = 0 \).
From \cref{eq:state}, the state at time \( \kappa\tau \) is
\begin{equation}
    \ket{\psi(\kappa\tau)}
    =
    \exp\bargs{-\i \phi \sigma^z} \ket+ ,
\end{equation}
where \( \phi = \int_0^{\kappa\tau} f(t')\gamma(t')\sigma^z \dd t' \).
We pick a measurement basis defined by \( \ket{\theta,\phi} = \cos\theta\ket+ +\e^{\i\phi}\sin\theta\ket- \) and any vector orthogonal to it.
We then consider
\begin{salign}
    \angles{P_{\theta,\phi}(\kappa\tau)}
    &= \angles{\abs{\bra{\theta,\phi}\ket{\psi(\kappa\tau)}}^2} \\
    &= \angles{\abs{\cos\theta\cos\phi - \i \e^{-\i\phi} \sin\theta\sin\phi}^2} \\
    &= \cos^2\theta \angles{\cos^2\phi} + \sin^2\theta \angles{\sin^2\phi} - \sin\phi \cos\theta\sin\theta \angles{\sin(2\phi)} \\
    &= \sin^2\theta + \cos(2\theta) \angles{\cos^2\phi} - \sin\phi \cos\theta\sin\theta \angles{\sin(2\phi)}.
\end{salign}
We proceed as in \cref{eq:avg-P+} to find that
\begin{equation}
    \angles{P_{\theta,\phi}(\kappa\tau)} = \sin^2\theta + \cos(2\theta) \angles{P_+(\kappa\tau)},
\end{equation}
where \( P_+(t) \) is, as before, \( \abs{\bra+\ket{\psi(t)}}^2 \).
Repeating the calculation done at \cref{eq:fisher-information-def} and below,
we again find that the Fisher information does not vanish as \( \Delta\omega\to 0 \)
only if \( F(\omega_c,\kappa\tau) = 0 \).

If on the other hand, we had considered \( P = \abs{\bra{\theta,0}\exp[-\i \phi\sigma^z] \ket{\theta,0}}^2 \), then we find
\( \angles{P} = \sin^2(2\theta) + \cos^2(2\theta)P_+(\kappa\tau) \).
And again we can only achieve superresolution if \( F(\omega_c,\kappa\tau) = 0 \).

\section{Error analysis of superresolution protocols}
\label{ap:error-analysis}

\subsection{Noiseless error analysis}
\label{sec:noiseless-error-analysis}

Recall that for each realization of $A_1,A_2,B_1,B_2$ in \cref{eq:signal}, we sample from a binomial distribution specified by the probability $P$.
Averaging over all these realizations yields $\angles*{P} = a - b \Delta\omega^2 + c \Delta\omega^4$, where we will drop all $\bigO*{\Delta\omega^6}$ terms.
From \cref{eq:probability-inst,eq::chi}, 
\begin{salign}
&a = \frac{1}{2}[1+e^{-4g^2F(\omega_c,T)}],\\
&b = \frac{g^2}{4}e^{-4g^2F(\omega_c,T)} F''(\omega_c,T),
\end{salign}
and $c$ can be similarly expressed in terms of the filter function.
Performing the sampling $N$ times, we therefore have $N$ random variables $\psi_i \in \set*{0,1}$. 
Our estimator $\widetilde{\Delta\omega}$ of $\Delta\omega$ is then $ \widetilde{\Delta\omega} = \frac{1}{\sqrt{b}} \sqrt{a - \frac{1}{N}\sum_i \psi_i}$.
We will denote the expected value over the realizations and over the corresponding samples by $\Expval$.

We will use that
\begin{equation}
    \frac{\abs{\Expval \sqrt{X} - \sqrt{\Expval X}}}{\sqrt{\Expval X}} \leq \frac{\operatorname{Var} X}{2 (\Expval X)^2} = \frac{\Expval X^2 - (\Expval X)^2}{2 (\Expval X)^2}
\end{equation}
for any random variable $X$ \cite{121424}.
Setting $X = \widetilde{\Delta\omega}^2 = \frac{1}{b}(a - \frac{1}{N}\sum_i \psi_i)$, we have $\Expval X = \Delta\omega^2 - \frac{c}{b}\Delta\omega^4$ and therefore
\begin{salign}
    \frac{\abs{\Expval \widetilde{\Delta\omega} - \Delta\omega}}{\Delta\omega}
    &\leq \frac{\Expval X^2 - (\Delta\omega^2-\frac{c}{b}\Delta\omega^4)^2}{2 \Delta\omega^4} + \bigO{\Delta\omega^8} \\
    &= \frac{\frac{a^2}{b^2} - \frac{2a}{b^2N} \sum_i \Expval \psi_i + \frac{1}{b^2 N^2}\sum_{i,j} \Expval \psi_i\psi_j - \Delta\omega^4}{2 \Delta\omega^4} + \bigO{\Delta\omega^8} \\
    &= \frac{\frac{a^2}{b^2} - \frac{2a}{b^2}(a-b\Delta\omega^2+c\Delta\omega^4) +  \frac{1}{b^2 N^2}\sum_{i} \Expval \psi_i^2 + \frac{1}{b^2 N^2}\sum_{i\neq j} \Expval \psi_i\psi_j - \Delta\omega^4}{2 \Delta\omega^4} + \bigO{\Delta\omega^2} \\
    &= \frac{\frac{a^2}{b^2} - \frac{2a}{b^2}(a-b\Delta\omega^2+c\Delta\omega^4) + \frac{1}{b^2 N^2}\sum_{i} \Expval \psi_i + \frac{1}{b^2 N^2}\sum_{i\neq j} \Expval \psi_i\psi_j - \Delta\omega^4}{2 \Delta\omega^4} + \bigO{\Delta\omega^2} \\
    &= \frac{\frac{a^2}{b^2} - \frac{2a}{b^2}(a-b\Delta\omega^2+c\Delta\omega^4) + \frac{a-b\Delta\omega^2+c\Delta\omega^4}{b^2 N} + \frac{N(N-1)(a-b\Delta\omega^2+c\Delta\omega^4)^2}{b^2 N^2} - \Delta\omega^4}{2 \Delta\omega^4} + \bigO{\Delta\omega^2} \\
    &= \frac{a(1-a)}{2b^2 N \Delta\omega^4} + \frac{2a-1}{2 b N \Delta\omega^2} + \frac{c-2ac-b^2}{2b^2 N} + \bigO{\Delta\omega^2} \\
    &\leq \frac{a(1-a)}{2b^2 N \Delta\omega^4} + \frac{2a-1}{2 b N \Delta\omega^2} + \frac{\abs{c}(1+2a)}{2b^2 N} + \bigO{\Delta\omega^2} 
    \eqqcolon \frac{B}{\Delta\omega}.
\end{salign}
It follows that in the single qubit superresolution case where we measure $N \sim \frac{1}{b \delta^2 \Delta\omega^2}$ times and $a = 1$, the relative bias is at most $\sim\delta$.

Let's consider the superresolution case, and denote the bias bound $B = \frac{1}{2 b N \Delta\omega} + \frac{3\abs{c}\Delta\omega}{2b^2 N}$, and the true bias $\calB$, so that $0 \leq \abs{\calB} \leq B$.
Note that $\calB = \Delta\omega - \Expval \widetilde{\Delta\omega} \geq 0$ by Jensen's inequality, so that $0\leq \calB \leq B$.
We have
\begin{salign}
    \Pr{}{\abs{\widetilde{\Delta\omega} - \Delta \omega} > \delta \Delta\omega} 
    &\leq \Pr{}{\widetilde{\Delta\omega} - \Delta \omega > \delta \Delta\omega}  + \Pr{}{\Delta \omega - \widetilde{\Delta\omega} > \delta \Delta\omega} \\
    &= \Pr{}{\widetilde{\Delta\omega} - \calB - \Expval \widetilde{\Delta\omega} > \delta \Delta\omega}  + \Pr{}{\calB + \Expval \widetilde{\Delta\omega} - \widetilde{\Delta\omega} > \delta \Delta\omega} \\
    &= \Pr{}{\widetilde{\Delta\omega} - \Expval \widetilde{\Delta\omega} > \delta \Delta\omega + \calB}  + \Pr{}{\Expval \widetilde{\Delta\omega} - \widetilde{\Delta\omega} > \delta \Delta\omega - \calB} \\
    &\leq \Pr{}{\widetilde{\Delta\omega} - \Expval \widetilde{\Delta\omega} > \delta \Delta\omega}  + \Pr{}{\Expval \widetilde{\Delta\omega} - \widetilde{\Delta\omega} > \delta \Delta\omega - B} \\
    &\leq 2\Pr{}{\abs{\widetilde{\Delta\omega} - \Expval \widetilde{\Delta\omega}} > \delta \Delta\omega - B} \\
    (\text{Chebyshev's inequality})
    &\leq \frac{2\operatorname{Var}\widetilde{\Delta\omega}}{(\delta \Delta\omega - B)^2} \\
    &= \frac{2}{(\delta \Delta\omega - B)^2}
    \parentheses{
    \Expval \widetilde{\Delta\omega}^2 - (\Expval \widetilde{\Delta\omega})^2
    } \\
    &= \frac{2}{(\delta \Delta\omega - B)^2}
    \parentheses{
    \Delta\omega^2 - \frac{c}{b}\Delta\omega^4 - (\Delta\omega - \calB)^2
    }\\
    &= \frac{4\calB \Delta\omega - 2\calB^2 - 2\frac{c}{b}\Delta\omega^4}{(\delta \Delta\omega - B)^2} \\
    \label{eq:p}
    &\leq \frac{4B \Delta\omega + 2\frac{\abs{c}}{b}\Delta\omega^4}{(\delta \Delta\omega - B)^2},
\end{salign}
where note the additional requirement for this bound to hold is that $\delta\Delta\omega > B$, giving that
\begin{equation}
    \label{eq:nu-meas-bound}
    N > \frac{1}{2b \delta \Delta\omega^2} + \frac{3\abs{c}}{2b^2 \delta}.
\end{equation}
Suppose we set $N = \alpha \Delta\omega^{-\gamma}$, then this required bound turns into $\gamma \geq 2$ and $\alpha \geq \frac{\Delta\omega^{\gamma-2}}{2b\delta} + \bigO{\Delta\omega^\gamma}$. Fixing $\gamma = 2$ and dropping subleading terms, we find
\begin{equation}
    \Pr{}{\abs{\widetilde{\Delta\omega} - \Delta \omega} > \delta \Delta\omega}
    \leq  \frac{8 b \alpha}{(2b\alpha \delta - 1)^2},\qquad \text{if } \alpha > \frac{1}{2b\delta}.
\end{equation}
Thus, in order for our estimate to achieve a relative error of at most $\delta$ with probability $1-p$, we require
\begin{equation}
    N > \frac{1}{\Delta\omega^2} \max\cbargs{
        \frac{2}{bp \delta ^2}+\frac{3}{2 b\delta },
        \frac{1}{2b\delta}
    } \sim \frac{1}{bp \delta ^2 \Delta\omega^2}.
\end{equation}

\subsubsection{Entanglement advantage}
\label{ap:entanglement-advantage}

In our single qubit superresolution protocols, we initialized the state as $\ket+$, evolved with the signal Hamiltonian, and measured the POVM $\set{\ket+\bra+, \bbI - \ket+\bra+}$.
For an entangled protocol using $N_e$ entangled qubits, we initialize in the GHZ state $\ket{\psi_0} \propto \ket{0}^{\otimes N_e} + \ket{1}^{\otimes N_e}$, evolve with the signal Hamiltonian separately on each qubit (we assume that the Hamiltonian on each qubit has the same realization of $A_1,A_2,B_1,B_2$ in \cref{eq:signal}), and measure the POVM $\set{\ket{\psi_0}\bra{\psi_0}, \bbI - \ket{\psi_0}\bra{\psi_0}}$.
The resulting measurement is again a binomial random variable. The probability defining this binomial distribution is the same as the single qubit case, except with $b\mapsto N_e^2 b$.
Thus, after $N_r$ repetitions of the experiment, the analysis from the previous section remains the same, except that $b \mapsto N_e^2 b$ and $N \mapsto N_r$.

Let $N_e = \beta / \Delta\omega$.
\cref{eq:p} and \cref{eq:nu-meas-bound} become
\begin{salign}
    &N_r > \frac{1}{2b \beta^2 \delta} + \bigO{\Delta\omega^4} \\
    &p = \frac{8b N_r \beta^2}{(1-2b N_r \beta^2 \delta)^2}.
\end{salign}
Note however that we have implicitly been assuming the small $\Delta\omega$ Taylor expansion of the probability $\angles{P} = 1 - b\Delta\omega^2$, so that in the entangled case we have $\angles{P} = 1 - b N_e^2\Delta\omega^2$.
However, this expansion breaks down when $bN_e^2 \Delta\omega^2 \approx 1$.
Namely, the $c$ in the expansion $\angles*{1-b\Delta\omega^2 + c \Delta\omega^4}$ generally has magnitude $\sim b^2$. Therefore, with the assigment $b\to N_e^2 b$, \cref{eq:nu-meas-bound} implies that our bounds are meaningful only when $\frac{1}{b\Delta\omega^2}\gtrsim 1$.
For $N_e = \beta / \Delta\omega$, it follows that it will break down when $b\beta^2 \sim 1$.
Setting $\beta^2b=1$, we see that we will always need at least $\sim1/\delta$ repetitions in order the achieve accuracy with constant probability.
To rigorously achieve superresolution, we should in fact set $N_e = \beta / \Delta\omega^{1-\mu}$ for $\mu > 0$.
The result is then that 
\begin{equation}
    \Pr{}{\abs{\widetilde{\Delta\omega} - \Delta \omega} > \delta \Delta\omega}
    \leq 
    \frac{8\beta^2 b N_r \Delta\omega^{2\mu}}{(1-2b \beta^2 N_r \delta \Delta\omega^{2\mu})^2},
\end{equation}
and therefore, the number of repetitions must grow as 
\begin{equation}
    N_r \gtrsim \frac{1}{\beta^2 b p \delta^2 \Delta\omega^{2\mu}}.
\end{equation}
We therefore see that even in the entangled protocol, no matter the number of entangled qubits, the number of repetitions must grow as $\Delta\omega\to 0$, though it can grow very slowly.

To compare the unentangled and entangled protocols, we compare the number of required measurements $N$ in the unentangled protocol to $N_e N_r$ in the entangled protocol. 
The entangled protocol uses less resources by a factor of $\frac{N}{N_e N_r} \sim \frac{\beta}{\Delta\omega^{1-\mu}} \sim N_e$.

\subsection{White Noise}
\label{ap:error-analysis-white-noise}

Let us consider the case of white noise, where $S(\omega)=\gamma$. Note, that we are working  in an idealized limit, since the noise has infinite power. Using the fact that the noise spectrum is independent of $\omega$, we can use the normalization of the filter function to show that the decay function is independent of any applied control,
\begin{align}
    \chi_\lambda(T=\kappa\tau)=2\gamma \kappa \tau
\end{align}

Recall that \( \angles{P} = \frac{1}{2} + \frac{1}{2}\e^{-\chi(\kappa \tau) - \chi_\lambda(\kappa \tau)} \).
For this analysis, we assume direct access to the probability distribution \( \angles*{P} \) because, as shown in \cref{sec:noiseless-error-analysis}, sampling from realizations of \( P \) rather than \( \angles*{P} \) introduces negligible error for superresolution protocols.
Assuming that the noise is weak, we can Taylor expand to leading order,
\begin{equation}
    P
    = 1- \frac{\alpha \pi^2g^2\kappa^2 \Delta\omega^2}{\omega_c^4} - \frac{4\pi \gamma\kappa }{\omega_c}+ \bigO{\Delta\omega^4},
\end{equation}
where \( \alpha^{\rm(free)} = 2 \) and \( \alpha^{\rm(CPMG)} = 8 \). 

Assuming that the estimation procedure utilizes the expression of noiseless evolution, the resulting estimate has a bias in addition to a variance. Explicitly, we sample \( N \) times from the binomial distribution defined by \( P \) in order to get an estimate $\tilde P$ of \( P \), and we estimate \( \Delta\omega \) as
\( \widetilde{\Delta\omega} = \sqrt{\frac{1-\tilde P}{\alpha\pi^2g^2\kappa^2/\omega_c^4}} \). 
The bias is,
\begin{equation}
    \label{eq:bias-white}
    b(\Delta\omega) \coloneqq \sqrt{\frac{1-P}{\alpha\pi^2g^2\kappa^2/\omega_c^4}} - \Delta\omega
    \qquad
    \implies \frac{b(\Delta\omega)}{\Delta\omega} =
    \sqrt{1 + \calW} - 1 + \bigO{\Delta\omega^2},
\end{equation}
where we defined $\calW \coloneqq \frac{4 \gamma \omega_c^3}{\alpha\pi g^2\kappa\Delta\omega^2}$.

Next, we bound the variance using \( \variance f(P) \leq (f'(\Expval P))^2 \variance P \), and that the variance of the binomial distribution after \( N \) measurements is \( P(1-P)/N \),
\begin{align}
    \variance{\widetilde{\Delta \omega}} \leq  \frac{1}{4(1-P)\alpha\pi^2g^2\kappa^2/\omega_c^4}\frac{P(1-P)}{N}= \frac{P}{4N\alpha\pi^2g^2\kappa^2/\omega_c^4}\leq \frac{1}{4 N \alpha \pi^2 g^2 \kappa^2 / \omega_c^4}
\end{align}

Then, using Chebyshev's inequality, 
\begin{equation}
    \Pr{}{\abs{\widetilde{\Delta\omega} - \Delta\omega} > b(\Delta\omega) + \varepsilon}
    = \Pr{}{\abs{\widetilde{\Delta\omega} - \Delta\omega - b(\Delta\omega)} > \varepsilon} 
    \leq \frac{\variance \widetilde{\Delta\omega}}{\varepsilon^2} 
    \leq \frac{1}{4 N \alpha \pi^2 g^2 \kappa^2 \varepsilon^2 / \omega_c^4},
\end{equation}
Hence, with probability at least \( 1-p \), our estimate of \( \Delta\omega \) will be within \( b(\Delta\omega) + \frac{1}{2\sqrt{p \alpha N} \pi g \kappa / \omega_c^2} \) of \( \Delta\omega \).

In order to compare the different protocols on equal footing, we always use \( N = N_{\rm free} = \frac{\omega_c^4}{8\pi^2g^2\kappa^2 \delta^2 \Delta\omega^2}\) measurements, that achieves a desired relative error $\delta$ for FE-SR.
Using this results in, with probability at least \( 1-p \), the relative error bound
\begin{equation}
    \frac{\text{Error}}{\Delta\omega} \leq \frac{b(\Delta\omega)}{\Delta\omega} + \frac{\delta}{\sqrt{p \alpha / 2}}
    = \sqrt{1+\calW} - 1 + \frac{\delta}{\sqrt{p \alpha / 2}} + \bigO{\Delta\omega^2}.
\end{equation}

\subsection{Lorentzian Noise}
\label{ap:error-analysis-lorentzian-noise}

In this section, we fix the noise \( \lambda(t) \) to have a Lorentzian spectrum as in \cref{eq:lorentzian-noise} with strength \( g_\lambda \) and full width at half maximum (FWHM) \( W \).
The corresponding contribution of this noise is
$\chi_\lambda(t) = \frac{1}{\pi} \int S_\lambda(\omega) F(\omega, t) \dd\omega.$
This section also serves as a rigorous analysis of the noise-free case by setting \( g_\lambda = 0 \). In this case, the decay factors are dependent on the applied control.
We consider the free \( \chi_\lambda^{(\rm free)} \) and CPMG \( \chi_\lambda^{(\rm CPMG)} \) protocols.
Define \( \tilde W \coloneqq \pi W / \omega_c \).
Using a sequence of change of variables, we find that
\begin{equation}
    \chi_\lambda(\kappa\tau) = \frac{8 \tilde W g_\lambda^2}{\omega_c^2} f_{\kappa,\tilde W},
\end{equation}
where
\begin{salign}
    f_{\kappa, \tilde W}^{\rm (free)}
    &= \int_{-\infty }^{\infty } \frac{\sin ^2(\kappa  \omega )}{\omega ^2 (4 \omega ^2 + \tilde W^2)} \dd\omega
    \leq \frac{\pi\kappa^2}{2\tilde W} \\
    f_{\kappa, \tilde W}^{\rm (CPMG)}
    &= 4\int_{-\infty }^{\infty } \frac{\sin ^4(\omega /2) \sec ^2(\omega ) \sin ^2(\kappa  \omega )}{\omega ^2 (4 \omega ^2 + \tilde W^2)} \dd\omega
    \leq f_{\kappa, 0}^{\rm(CPMG)}.
\end{salign}
Note that \( f_{\kappa, \tilde W}^{\rm (free)} \) goes to \( \infty \) as \( \tilde W \to 0 \), whereas \( f_{\kappa,\tilde W}^{\rm (CPMG)} \) does not. For example, \( f_{2, \tilde W}^{\rm(CPMG)} \leq f_{2, 0}^{\rm(CPMG)} = \pi/6 \), \( f_{4,0}^{\rm(CPMG)} = \pi/3 \), and \( f_{6,0}^{\rm(CPMG)} = \pi/2 \).

Recall that \( \angles{P} = \frac{1}{2} + \frac{1}{2}\e^{-\chi(\kappa \tau) - \chi_\lambda(\kappa \tau)} \).
For this analysis, we assume direct access to the probability distribution \( \angles*{P} \) because, as shown in \cref{sec:noiseless-error-analysis}, sampling from realizations of \( P \) rather than \( \angles*{P} \) introduces negligable error for superresolution protocols.
We can easily check that
\begin{equation}
    P
    = 1- \frac{\alpha \pi^2g^2\kappa^2 \Delta\omega^2}{\omega_c^4} - \frac{4g_\lambda^2 \tilde W}{\omega_c^2} f_{\kappa, \tilde W} + \bigO{\Delta\omega^4}.
\end{equation}
where \( \alpha^{\rm(free)} = 2 \) and \( \alpha^{\rm(CPMG)} = 8 \).

In our superresolution protocols, we sample \( N \) times from the binomial distribution defined by \( P \) in order to estimate \( P \), and we estimate \( \Delta\omega \) as
\( \widetilde{\Delta\omega} = \sqrt{\frac{1-P}{\alpha\pi^2g^2\kappa^2/\omega_c^4}} \). The resulting bias is
\begin{equation}
    \label{eq:bias}
    b(\Delta\omega) \coloneqq \sqrt{\frac{1-P}{\alpha\pi^2g^2\kappa^2/\omega_c^4}} - \Delta\omega
    \qquad
    \implies \frac{b(\Delta\omega)}{\Delta\omega} =
    \sqrt{1 + \calW} - 1 + \bigO{\Delta\omega^2},
\end{equation}
where we defined
\begin{equation}
    r_{g_\lambda} \coloneqq \sqrt{\frac{1}{\pi}} \cdot \frac{g_\lambda}{g} \cdot \frac{\omega_c}{\Delta \omega}, \qquad
    \calW \coloneqq \frac{4 f_{\kappa, \tilde W} \kappa  r_{g_\lambda}^2 \tilde W}{\pi  \alpha \kappa^2}.
\end{equation}
Then, using Chebyshev's inequality,
\begin{salign}[eq:Chebyshev-lorentz]
    \Pr{}{\abs{\widetilde{\Delta\omega} - \Delta\omega} > b(\Delta\omega) + \varepsilon}
    &= \Pr{}{\abs{\widetilde{\Delta\omega} - \Delta\omega - b(\Delta\omega)} > \varepsilon} \\
    &\leq \frac{\variance \widetilde{\Delta\omega}}{\varepsilon^2} \\
    &\leq \frac{P (1-P)/N}{4 \alpha \pi^2 g^2 \kappa^2 \varepsilon^2 (1-P) / \omega_c^4} \\
    &\leq \frac{1}{4 N \alpha \pi^2 g^2 \kappa^2 \varepsilon^2 / \omega_c^4},
\end{salign}
where we used that
the variance of the binomial distribution after \( N \) measurements is \( P(1-P)/N \) and that
\( \variance f(P) \leq (f'(\Expval P))^2 \variance P \).
Hence, with probability at least \( 1-p \), our estimate of \( \Delta\omega \) will be within \( b(\Delta\omega) + \frac{1}{2\sqrt{p \alpha N} \pi g \kappa / \omega_c^2} \) of \( \Delta\omega \).

In \cref{fig:MSE-numerics,fig:cc-mse}, in order to compare the different protocols on equal footing, we always use \( N = N_{\rm free} = \frac{\omega_c^4}{8\pi^2g^2\kappa^2 \delta^2 \Delta\omega^2}\) measurements.
Using this results in, with probability at least \( 1-p \), the relative error bound
\begin{equation}
    \frac{\text{Error}}{\Delta\omega} \leq \frac{b(\Delta\omega)}{\Delta\omega} + \frac{\delta}{\sqrt{p \alpha / 2}}
    = \sqrt{1+\calW} - 1 + \frac{\delta}{\sqrt{p \alpha / 2}} + \bigO{\Delta\omega^2}.
\end{equation}
Plugging in the values for \( \calW \) and \( \alpha \) yields
\begin{salign}[eq:analytic-error-bounds]
    \frac{\text{Error}^{\rm(free)}}{\Delta\omega}
    &\leq
    \sqrt{1+r^2_{g_\lambda}} - 1 + \frac{\delta}{\sqrt{p}}
    + \bigO{\Delta\omega^2}
    \\
    \frac{\text{Error}^{\rm(CPMG)}}{\Delta\omega}
    &\leq
    \sqrt{1+r_{g_\lambda}^2 \cdot \frac{W f_{\kappa, 0}}{2\kappa^2 \omega_c}} - 1 + \frac{\delta}{2\sqrt{p}}
    + \bigO{\Delta\omega^2}.
\end{salign}
To recover the error bounds for the noise-free case, \( r_{g_\lambda} \) and \( W \) can be set to \( 0 \).
We note that the \( \bigO{\Delta\omega^2} \) terms become important for large \( \Delta\omega \); the resulting error is due to the Taylor series estimation no longer being a valid approximation.

As \( \Delta\omega\to 0 \) and when \( \kappa = 2 \), we see that CPMG outperforms free evolution by a factor of more than \( \sqrt{15 \omega_c/W} \). This result is very intuitive. In particular, CPMG outperforms free evolution because of the FWHM \( W \), not because of the strength \( g_\lambda \).
The smaller \( W / \omega_c \) is, the less overlap the CPMG filter function has with the noise.
The effect of \( g_\lambda \) on the error shows up in the same way in both CPMG and free evolution.
When \( \omega_c = 1 \) and \( W = 0.1 \), we see that CPMG should achieve a lower relative error than free evolution by a factor of more than \( \approx 12  \) as \( \Delta\omega \to 0 \), which we indeed see in the numerical simulations presented in \cref{fig:MSE-numerics}.

\subsubsection{Subtracting off the bias}
\label{sec:subtract-bias}

Suppose we somehow know (even \textit{exactly}) the noise strength $g_\lambda$ and FWHM $W$.
Can we then subtract off the noise and regain perfect superresolution even as \( \Delta\omega\to 0 \)?
The answer is no, as we show now.

Given that we know the noise, our estimator $\widetilde{\Delta\omega}$ becomes $\sqrt{\frac{1-P - 8g_\lambda^2 \tilde W f_{\kappa,\tilde W}/\omega_c^2}{\alpha \pi^2 g^2 \kappa^2 / \omega_c^4}}$.
It follows that the bias $b(\Delta\omega)$ goes to zero as $\Delta\omega \to 0$.
Then, proceeding as in \cref{eq:Chebyshev-lorentz} and working only to order $\Delta\omega^2$, we have
\begin{salign}
    \Pr{}{\abs{\widetilde{\Delta\omega} - \Delta\omega} > \varepsilon}
    &\leq \frac{\variance \widetilde{\Delta\omega}}{\varepsilon^2} \\
    &\leq \frac{P (1-P)/N}{4 \alpha \pi^2 g^2\kappa^2 \varepsilon^2 (1-P - 8g_\lambda^2 \tilde W f_{\kappa,\tilde W}/\omega_c^2) / \omega_c^4} \\
    &\leq \frac{\frac{\alpha \pi^2g^2\kappa^2 \Delta\omega^2}{\omega_c^4} + \frac{8g_\lambda^2 \tilde W}{\omega_c^2} f_{\kappa, \tilde W}}{4 N \alpha \pi^2 g^2\kappa^2 \varepsilon^2 (\frac{\alpha \pi^2g^2\kappa^2 \Delta\omega^2}{\omega_c^4}) / \omega_c^4}.
\end{salign}
It follows that in order to estimate $\Delta\omega$ to a relative error of $\delta = \varepsilon / \Delta\omega$ with high probability as $\Delta\omega\to 0$, we need
$N \sim \frac{1}{\Delta\omega^4}$.

Thus, we see that even if we knew the noise exactly, subtracting off the noise gets rid of the bias but does not strictly allow us to regain superresolution.
This comes down to sampling from the binomial distribution defined by $P = a - b \Delta\omega^2$, as discussed in the main text. When $a \notin \{0, 1 \}$, the variance of the binomial distribution prevents superresolution.


\twocolumngrid
\bibliographystyle{apsrev4-2}
\bibliography{references}

\end{document}